\documentclass[twocolumn]{aastex631}

\newcommand{\ltsima}{$\; \buildrel < \over \sim \;$}
\newcommand{\simlt}{\lower.5ex\hbox{\ltsima}}
\newcommand{\gtsima}{$\; \buildrel > \over \sim \;$}
\newcommand{\simgt}{\lower.5ex\hbox{\gtsima}}

\newcommand{\boo}{{Boo~I}}

\graphicspath{{./}{figures/}}
\received{August 18, 2021}
\revised{October 4, 2021}
\accepted{October 6, 2021}
\submitjournal{{ApJ}}

\shorttitle{Wide-area view of \boo}
\shortauthors{Filion \& Wyse}

\begin{document}

\title{The Far-Away Blues: Exploring the Furthest Extents of the Bo{\"o}tes~I Ultra Faint Dwarf Galaxy}

\correspondingauthor{Carrie Filion}\email{cfilion@jhu.edu}
\author[0000-0001-5522-5029]{Carrie Filion}
\affil{{Department of Physics \& Astronomy, The} Johns Hopkins University, {Baltimore, MD 21218}}

\author[0000-0002-4013-1799]{Rosemary F.G.~Wyse}
\affiliation{{Department of Physics \& Astronomy, The} Johns Hopkins University, {Baltimore, MD 21218}}



\begin{abstract}
Establishing the spatial extents and the nature of the outer stellar populations of  dwarf galaxies are necessary ingredients for the determination of their total masses, current dynamical states and past  evolution.  We here describe our investigation of the outer stellar content of the Bo{\"o}tes~I ultra-faint dwarf galaxy,  a satellite of the Milky Way Galaxy. We identify candidate member blue horizontal branch and blue straggler stars of Bo{\"o}tes~I, both tracers of the underlying ancient stellar population, using a combination of multi-band Pan-STARRS photometry and \textit{Gaia} astrometry. We find a total of twenty-four candidate blue horizontal branch member stars with apparent magnitudes and proper motions consistent with membership of Bo{\"o}tes I, nine of which reside at projected distances beyond the nominal King profile tidal radius derived from earlier fits to photometry. We also identify four blue straggler stars of appropriate apparent magnitude to be at the distance of Bo{\"o}tes I, but all four are too faint to have high-quality astrometry from \textit{Gaia}. The outer blue horizontal branch stars that we have identified confirm that the spatial distribution of the stellar population of Bo{\"o}tes~I is quite extended. The morphology on the sky of these outer envelope candidate member stars is evocative of tidal interactions, a possibility that we explore further with simple dynamical models.
\end{abstract}



\section{The Stellar Content of the Bo{\"o}tes~I Ultra-Faint Dwarf Galaxy} \label{sec:intro}

Ultra-faint dwarf (UFD) galaxies were discovered over the last fifteen or so years, with the advent of all-sky imaging surveys  \citep[see][for a recent review]{simon-review}.  They  may be survivors from the earliest  stages of the formation of structure in the Universe \citep[e.g.][]{bovill_2009} and as such, hold unique clues as to how stars and galaxies form. Establishing their spatial extents and the nature of their outer stellar populations  are necessary ingredients for the determination of their total masses, current dynamical states and past  evolution.  As recently discussed by \citet{chiti}, discovery and characterization of an extended stellar envelope provides important insight into the past merger history of the host dark-matter halo, and UFDs probe the mass regime in which the standard Cold Dark Matter paradigm faces challenges \citep{bullock}.  We here describe our investigation of the outer stellar content of the Bo{\"o}tes~I ultra-faint dwarf galaxy (hereafter Boo~I), a satellite of the Milky Way Galaxy.

\boo\ was discovered by \citet{belokurov06} in the imaging data from the Sloan Digital Sky Survey. It is relatively luminous for a UFD (M$_V \sim -6$; $L_V \sim 2 \times 10^4 L_\odot$; \citep{belokurov06}) and nearby, with a heliocentric distance of $\sim 65$~kpc (\citealp[e.g.][]{okamoto}), and has been the target of several spectroscopic and photometric surveys. The internal stellar velocity dispersion of \boo\ is sufficiently high ($\sim 5$~km/s) that, in combination with the half-light radius ($\sim 200$~pc), dark matter dominates the mass \citep[see][for a recent analysis]{jenkins}, as is typical for a UFD.  The isophotes tracing the stellar surface density are flattened and irregular, suggestive of the effects of tidal interactions with the Milky Way (see \citealp{belokurov06,fellhauer,roderick}). We focus on quantifying the extent of the outer envelope of stars, using a new technique to identify blue stars that are likely to be members.

The color-magnitude diagram (CMD) of \boo\  matches that of  an ancient, metal-poor stellar population, containing stars with apparent magnitudes and colors consistent with being    blue horizontal branch (BHB) stars, together with a population of candidate blue-straggler stars (BSS), brighter than the old main-sequence turn-off \citep{belokurov06, okamoto}. The blue stragglers follow a similar radial profile to the red-giant branch stars, at least out to two half-light radii \citep{santana}, supporting that BSS formed via mass transfer in close-binary systems belonging to the underlying old population.   Thus both BHB and BSS trace the dominant old, metal-poor population \citep[see also][]{momany}.  Spectroscopy of bright RGB stars confirmed the low metallicity, with a mean below $-2$~dex  \citep{munoz06,martinspec,norris08} and identified metal-poor, radial-velocity members at several half-light radii,  with a hint of a radial metallicity gradient \citep{norris08, norris_2010b}.  Indeed, some of these stars lie beyond the reported nominal tidal radius, obtained from a King-model fit to the stellar light profile.\footnote{We adopt the value of 37.5~arcmin for the King profile tidal radius, found by \citet{munoz}, which closely matches that of \citet{okamoto},  is consistent with that of \citet{roderick}, and is the largest of the three estimates.} We note that the physical interpretation of such a tidal radius, obtained from a model developed for self-gravitating star clusters, is not clear for dark-matter dominated systems (see, for example, the discussion in \citealt{read}). However, this photometric tidal radius does indicate the (model-dependent) edge of the stellar surface-brightness profile and is frequently used in the literature as a measure of the extent of the luminous matter in galaxies, in addition to star clusters. We will, henceforth, identify this photometric tidal radius as the ``King profile nominal tidal radius'', in the hope of avoiding confusion with a more dynamically motivated tidal radius that accounts for dark matter. We will refer to stars at projected distances beyond this radius as ``outer envelope stars". \citet[see their Fig.~3]{belokurov06} noted an  envelope of candidate BHB and BSS,  selected purely on their location in the observed color-magnitude diagram, extending in projected radius to the limits of their photometric data (a $1^{\rm o} \times 1^{\rm o}$ field). Similarly, \citet{roderick} identified candidate BHB and BSS - again defined purely on their location in the CMD -  to the radial limits of their 3~square~degree field; these authors further found that the BHB and BSS candidates followed the same radial profile, with a change in slope at around the half-light radius. There is significant irregularity and substructure in the outer regions of \boo, and it is unclear whether or not the (candidate) BHB and BSS follow the pronounced elongation of the higher surface density inner regions \citep{belokurov06,roderick}. Several BHB candidates  were included in the spectroscopic study of \citet{koposov} that targeted the central regions of Boo~I, and all (7/7) were found to have both radial velocity and proper motion (from {\it Gaia\/} eDR3) consistent with membership \citep{jenkins}. One of these BHB candidates was recognized as an RR~Lyrae variable previously identified by \citet{siegel} as a possible member (his V2).

Indeed, $15$ RR Lyrae variables with apparent magnitudes consistent with being at the distance of \boo\ were identified in two dedicated surveys for candidate member variable stars, covering a maximum 46~arcmin field  \citep{siegel,dallora}.  The vast majority (12) of these lie within the half-light radius, with one beyond the King profile nominal tidal radius. \citet{vivas} used the catalog of variable stars  from {\textit{Gaia} DR2 photometry, compiled by \citet{clementini}, to search for RR Lyrae within a 2~degree radius of the center of \boo. They recovered only 2 of those that had been previously identified, presumably reflecting the \textit{Gaia} scanning pattern at the time of DR2, and added one new candidate, which also lies  beyond the King profile nominal tidal radius.  As noted by \citet{siegel}, should they all be members, \boo\ would have an unusually high frequency of RR Lyrae, given its estimated total luminosity. \citet{vivas} showed that the \textit{Gaia} DR2 proper motions for all but one of the 16 RR Lyrae candidates are consistent with membership; the uncertainties were however large and we revisit  the RR Lyrae using proper motions from \textit{Gaia} eDR3 in Section~\ref{sec:mean-pm}  below.\footnote{\citet{siegel} also identified a bright variable star with an apparent magnitude close the tip of the red-giant branch in \boo, identified as a  long period variable star (LPV) by \citet{dallora}. Should this LPV be a member of \boo, it would likely be the lowest metallicity LPV known and we discuss its status further in Section~\ref{sec:LPVsec}.}

There are thus several lines of evidence pointing to an envelope of stars physically associated with \boo, essentially out to the limits of the available imaging datasets. The implications, in terms of the total mass and mass profile of \boo\ (both stellar and dark matter), together with possible mechanisms by which this envelope could have been populated, obviously depend on whether or not these stars are indeed members. We address this question here, taking advantage of the advent of public wide-area imaging surveys and the astrometry of {\it Gaia\/} eDR3. We first develop a technique to identify stars in the very outer parts of \boo, and then discuss their membership status.  We focus on blue stars, as for these stars the contamination by foreground stars in the Milky Way is lessened (though cannot be ignored).    The need to maximize the contrast with the foreground, especially when pushing further from the center of \boo, is illustrated in  Figure~\ref{fig:panstarrs_cmd}, which shows the color-magnitude diagrams derived from Pan-STARRS (PS1) photometry \citep{panstarrs} for two fields of different extent, both centered on \boo. The selection criteria for the photometry and the extinction-correction procedure are discussed below, in Section~\ref{sec:photometry_selection}. The lefthand panel shows all stellar sources within the half-light radius of \boo\ ($r_h = 10.5 $~arcmin, or $\sim 200$pc, adopting the structural parameters of the exponential fit from \citealp{munoz}), while the righthand panel shows all stellar sources within a four-by-four degree area. A PARSEC isochrone \citep{bressan} of age 13~Gyr and [M/H]$ =-2.2$ (the lowest metallicity available), shifted to the heliocentric distance of \boo\ (65~kpc, \citealp{okamoto}, consistent with the RR Lyrae-based distances of 62 $\pm$ 4 and 66 $\pm$ 3 kpc of \citealt{siegel} and \citealt{dallora}, respectively),  is overplotted on the data. The over-density of sources near the horizontal branch of the isochrone is evident on the lefthand panel, together with  a hint of possible BSS at fainter magnitudes. These features are less well-defined in the righthand panel, which also shows the dominance of the foreground at colors redder than the main-sequence turn-offs of the field stellar halo and thick disk, motivating our focus on blue stars in \boo. 

\begin{figure}
\begin{center}
\includegraphics[width=.45\textwidth]{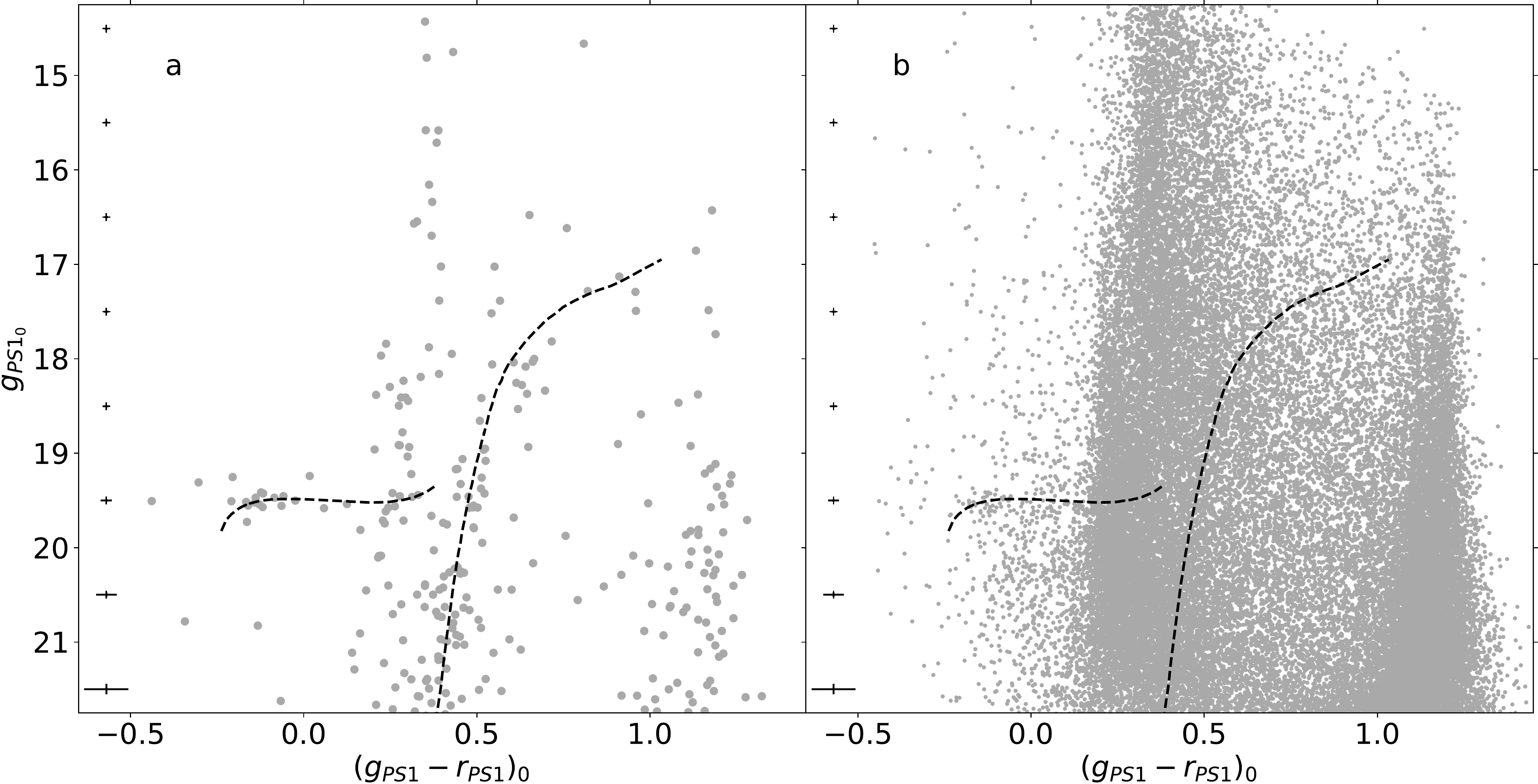}

\caption{The CMD of all PS1 sources that pass our quality control cuts  within (a) the half-light radius of the nominal center of \boo\ (lefthand panel), and (b)  a four-by-four degree area centered on \boo\ (righthand panel). The locus of a PARSEC \citep{bressan} isochrone of age 13~Gyr, with [M/H]$ = -2.2$, shifted by the distance modulus of \boo, is over-plotted in dashed black lines in both panels. The symbol size of sources in the righthand panel is reduced, relative to the lefthand panel, for clarity.}

\label{fig:panstarrs_cmd}
\end{center}
\end{figure}

After we completed the analysis presented below, the complementary analysis of \cite{longeard} was posted to the arXiv. These authors similarly  identify candidate \boo\ member stars far from that galaxy's center  using a combination of photometry and \textit{Gaia\/} astrometry. In their case they utilize the metallicity-sensitive  photometric filters from the Pristine survey \citep[see e.g.][]{pristine} and focus on RGB and HB stars redder than $(g-i)_{0,\, {\rm SDSS}} \sim 0.2$. There is no overlap between the sample of \cite{longeard} and the candidate member stars identified using our technique below.  We  discuss the results from \cite{longeard} further, within the context of our findings, in Section~\ref{sec:discussion}  below.

\vfill
\eject

\section{Photometric Identification of Candidate BHB and BSS Stars}\label{sec:photometry_selection}

Given a sample of blue stars, the identification of BHB stars, which have lower surface gravities than do main-sequence A-type stars of similar effective temperatures (and colors), through photometry has a long history and has evolved from UBV(RI) data \citep[e.g.,][]{newell,wilhelm} to some combination of $ugriz$ \citep[e.g.,][]{yanny,vickers}. The technique essentially combines a color that is temperature-sensitive with one that is gravity-sensitive (in this range of effective temperature). The U-band ($u$-band) filter covers the Balmer jump which is higher in lower-gravity A-type stars, while the $z$-band filter covers the Paschen jump, which is similarly gravity-sensitive.   We used photometry from the Pan-STARRS1 (PS1) survey, described in \citet{panstarrs}, which is well-suited to an analysis of the blue stellar content of \boo, despite its lack of $u$-band information, as it has sufficient depth to study the blue-straggler stars\footnote{The median five-sigma depths  in  $g,\, r,\, i,\/ {\rm  and}\, z$ are, respectively, 23.3, 23.2, 23.1 and 22.3 magnitudes, although the effective depth of our final catalog is shallower due to the selection criteria and quality cuts imposed.}   (as demonstrated in Figure~\ref{fig:panstarrs_cmd}) and includes the $z-$band, which, as just noted, can be used to discriminate among  A-type stars of differing surface gravities \citep{vickers}. We will use the location of a given blue star in $griz$ color-color space to identify candidate BHB and BSS stars in \boo. We did not utilize external $u$-band data, as we aimed to explore the utility of PS1 alone in the classification of blue sources. We compare the results of this $griz$-based analysis to those of a complementary $ugri$-based analysis in Appendix \ref{sec:knn_appendix}.

\subsection{Characterization of Different Stellar Sources in $griz$}
  
Even with a sample selected to consist of only blue stellar-like objects, there are several additional source types that can occupy the blue side of color-magnitude diagrams such as those in Fig.~\ref{fig:panstarrs_cmd}, potentially contaminating a sample of the BHB and BSS of interest. These include background sources such as high-redshift, star-forming galaxies, AGN and quasars. Foreground main-sequence A-stars are unlikely at the relevant apparent magnitude range, particularly at the Galactic coordinates of \boo\ ($\ell,\,b = 358^{\rm o},\, +70^{\rm o}$).  White dwarfs  are likely, but their extremely high surface gravities (and their parallaxes, if available) should provide a means to identify them.  Blue straggler stars and horizontal branch stars in the outer regions of the field halo of the Milky Way could also have colors and magnitudes  close to those of actual \boo\ members.

Guided by the earlier analyses of, for example, \citet{sirko,xue_2008,vickers}, we quantified the loci occupied by different types of blue star-like sources in color-color space by combining the photometry with information from spectroscopy.   We first selected all sources in the SDSS/SEGUE imaging data with `clean' photometry (i.e. flagged as such in the database) and observed SDSS color $ g-r \le 0.1$.  We then found the corresponding PS1 photometry for each source,\footnote{Specifically, we used the \texttt{PS1\_MeanObject} photometric catalog joined with  \texttt{PS1\_ObjectThin}, which provides positional information.} using a maximum matching radius of one~arc~second. We implemented quality cuts, following those recommended by \citet{flewelling}, namely that the quality-of-fit parameter \texttt{QfPerfect} be greater than 0.85 for each of the four filters of this study ($griz$) and that the number of detections be greater than 5. We removed sources that were  inconsistent with being stellar, based on their image shape and size compared to the point-spread function.\footnote{We required a maximum deviation from a point source in each filter, as measured by the difference between the two PS1-derived magnitudes  \texttt{MeanPSFMag}-\texttt{MeanKRONMag} $< 0.05$, as proposed by  \citet{farrow}.}   We further restricted the sample to  sources with extinction in the $g-$band less than or equal to 0.5~mag ($\sim 5\times$ the extinction along the line-of-sight towards \boo), based on the dust maps of  \citet{schlafly}, and extinction-corrected the photometry for the remaining 16,205 sources, using  the \texttt{dustmaps} package of \citet{green}. We will refer to this cross-matched sample as the SEGUE-PS1 photometric catalog; its depth is determined by the  SEGUE spectroscopic limit, at $g \sim 20$. 

\begin{figure}
\begin{center}
\includegraphics[width=.42\textwidth]{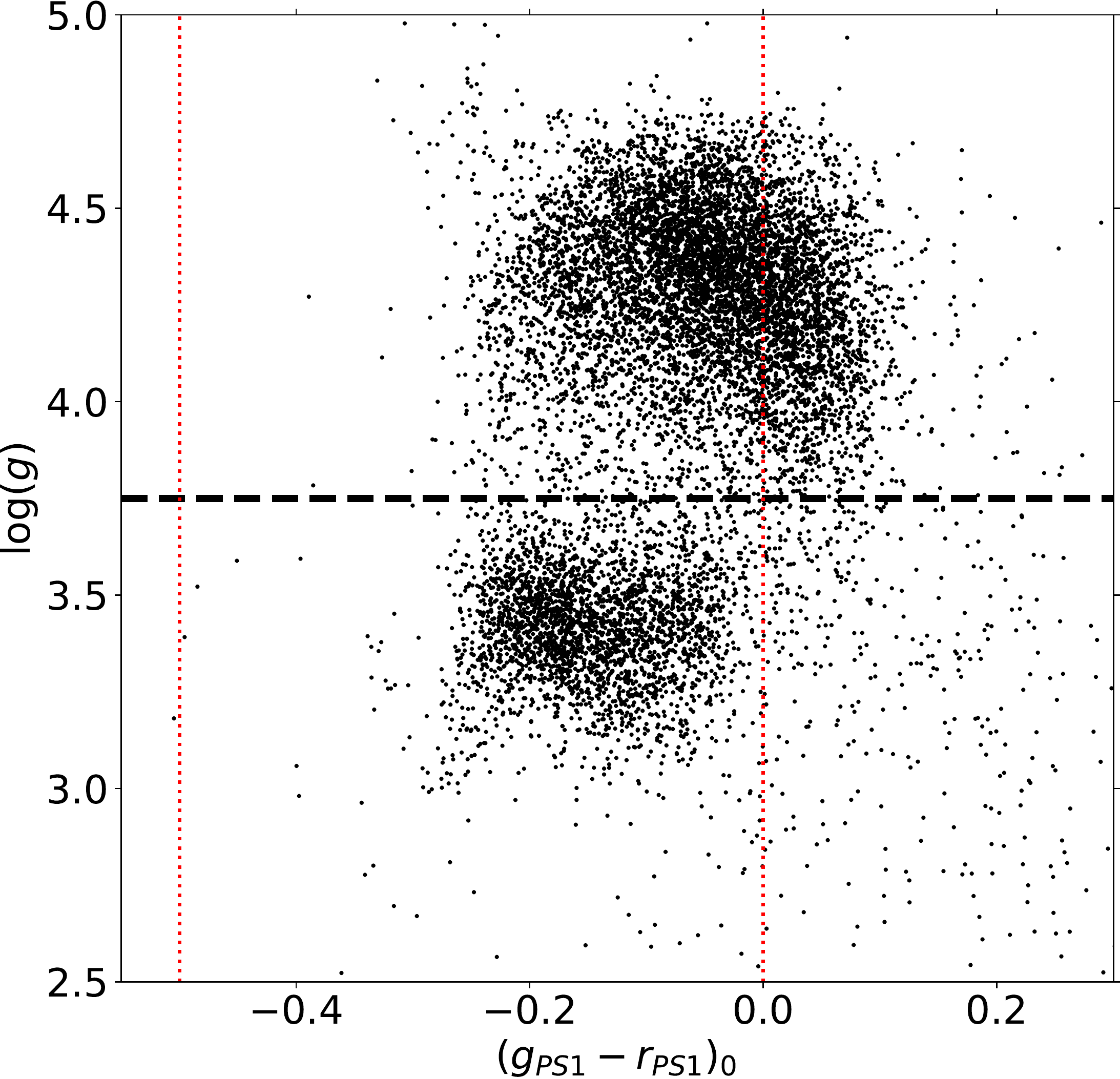}

\caption{A plot of $(g_{PS1}-r_{PS1})_0$ color vs $\log~g$ for the set of stellar sources with surface gravity estimates from the SEGUE/SSPP and contained in the  SEGUE-PS1 photometric catalog, derived as  described in the text. The thick, dashed line at $\log\, g = 3.75$ indicates  the boundary that we adopted to separate  BHB and BSS/MS stars. The thin, dotted, vertical red lines show the bluest and reddest extents of the color range adopted in the classification scheme developed in this paper.}

\label{fig:logg}
\end{center}
\end{figure}

The distribution  in the log~$g$ versus $(g_{PS1} - r_{PS1})_0$ color plane of the subset of sources from the SEGUE-PS1 photometric  catalog which also have estimates of  surface gravity from the SEGUE stellar parameter pipeline (SSPP) is shown in  Figure~\ref{fig:logg}.} There is a clear separation into two clumps, characterized by surface gravity. The lower surface gravity stars are consistent with being BHB and the higher gravity stars are consistent with being main-sequence A-type stars \citep[e.g.][]{bressan,vickers}. The horizontal thick, dashed line in Figure~\ref{fig:logg} at log~$g = 3.75$ indicates the dividing line in surface gravity that we used to assign blue stars to either of these two categories (the same threshold was adopted  by \citealp{vickers}). Specifically,  we labeled the higher-gravity blue stellar sources (log~$g > 3.75$) as BSS/MS and the lower-gravity ones  ($2.25 \le {\rm  log}\, g < 3.75$) as BHB.  White dwarfs do not have an estimated surface gravity in the SSPP, but are flagged as such (\lq D') in the SEGUE database. Similarly, sources with emission lines, and therefore  likely to be non-stellar, also lack an estimated surface gravity and are flagged (\lq E').  We  assigned the label `WD' to all sources flagged as likely WDs and  assigned the label  `QSO' to all flagged emission-line sources  (remembering that all sources in our SEGUE-PS1 photometric catalog have been required to have blue colors and stellar image profiles). All remaining sources, of which there were 4274, we labeled as ``other'';  11 are stars with gravity too low   (log~$g < 2.75$) to be low-mass core-helium-burning BHB  while the rest lack estimates for their atmospheric parameters for a variety of reasons. We did not consider any of these ``other'' sources in the analysis below.

\begin{figure*}
\begin{center}
\includegraphics[width=.9\textwidth]{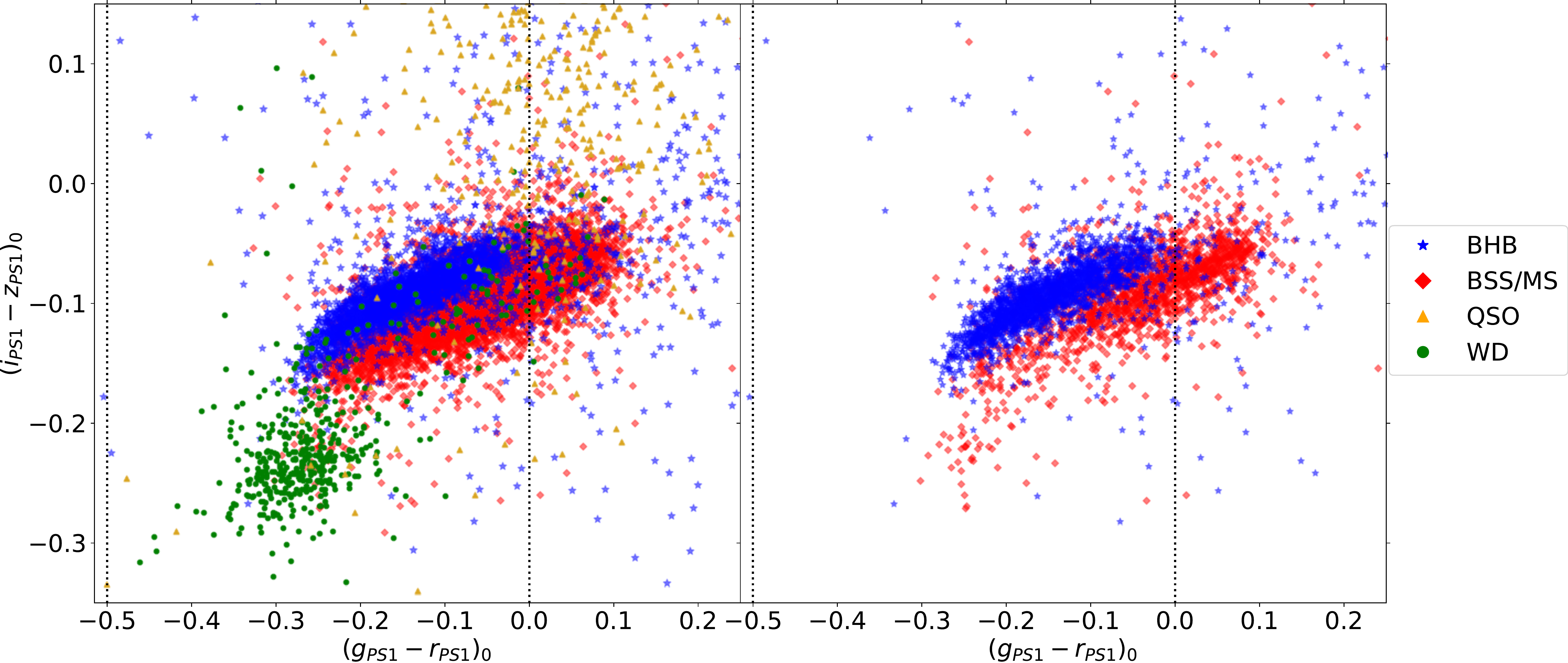}

\caption{A $(g_{PS1}-r_{PS1})_0$ vs $(i_{PS1}-z_{PS1})_0$ color-color diagram of star-like sources in the SEGUE-PS1 photometric catalog for which we derived labels. The blue stars represent BHB stars, red diamonds represent BSS/MS stars, green circles represent WDs and orange triangles represent QSOs. All labeled sources are shown in the lefthand panel while the stars in the righthand panel are limited to those  BHB and BSS/MS with metallicity estimates from the SSPP below $ -1.5$~dex. The vertical dotted lines indicate the bluest and reddest extents of the color range adopted in this study. }

\label{fig:segue_ccd}
\end{center}
\end{figure*}

The distribution of sources in the SEGUE-PS1 photometric catalog in the $(g_{PS1}-r_{PS1})_0$ versus $(i_{PS1}-z_{PS1})_0$ color-color diagram   is presented in Figure~\ref{fig:segue_ccd}, with each class identified by symbol type and color. The lefthand panel shows all sources we labeled as BHB, BSS/MS, WD, or QSO, while the righthand panel contains only the BHB and BSS/MS stars for which the SEGUE SSPP provided a low metallicity, specifically $\rm{[Fe/H]_{SSPP}} \le -1.5$,  approximately the maximum value found for stars in \boo\ with spectroscopic abundances.  While the metallicity distribution of stars in the metal-poor field halo, traced by the stars in the righthand panel, and that of \boo\ differ, a comparison of the two panels of Figure~\ref{fig:segue_ccd} shows that metallicity has little impact on the loci occupied by  the BHB and BSS/MS stars, affecting primarily the densities within the boundaries of those loci. We assumed below that this remains true even at extremely low values of the iron abundance, [Fe/H] $\sim -3$~dex, which is poorly represented within the  SEGUE/SSPP sample.

Armed with this characterization of these different classes of source in $griz$ space, we now turn to the application of this knowledge to samples which lack spectroscopic atmospheric parameters, such as the sources observed along the line-of-sight to \boo. As described below, we developed an automated classification technique that avoids subjective drawing of boxes in the color-color plane.       We made the further restriction to consider only sources with  colors in the range $-0.5 \le (g_{PS1}-r_{PS1})_0 \le 0$, which should  encompass the expected  BHB in \boo\  (see Figure~\ref{fig:panstarrs_cmd}), blueward of the instability strip.\footnote{The blue edge of the core-helium-burning instability strip in a metal-poor population occurs at $\log\, {\rm T}_{eff} \sim 3.85$   \citep[e.g., see Table~1 of][]{marconi}. We estimated the corresponding PS1 color, $(g_{PS1}-r_{PS1})_0 \sim 0$,  using the models of \citet{marconi} and the photometric conversions in \citet{tonry}.}

\vfill
\eject

\subsection{K-Nearest Neighbors Classification}\label{sec:knn-main}

As may be  seen in Figure~\ref{fig:segue_ccd}, it is not straightforward to separate  BHB from BSS/MS (hereafter BSS). We opted to use a k-nearest neighbors (KNN) classifier, which we trained on the SEGUE-PS1 photometric catalog. A KNN classifier acts first to identify the k objects (where k is some integer that the user is free to specify in order to optimize the process) in the training set that have the k smallest distances (in an appropriate sense)  to a given input object, and then to classify the input object based on the labels of those k-nearest objects, providing the probability of each possible label.

We used the \texttt{KNeighborsClassifier} implemented in the Python package scikit-learn \citep{sklearn}. We trained this classifier on subsets of the SEGUE-PS1 photometric catalog, using all possible color combinations of the PS1 $griz$ photometry, such that the classifier worked in six-dimensional color space. The KNN classifier was then tested on the subsets of the catalog that were not used in the training; details of the training and testing are given in Appendix~\ref{sec:knn_appendix}, with only a  brief summary here, outlining the procedure adopted and the success of the classifier.

The fractions of sources labeled as BSS, BHB, WD or QSO in the SEGUE-PS1 catalog are quite unequal, which could impact the classifier, so we first re-sampled the catalog to achieve a better balance between the classes (as described in more depth in Appendix~\ref{sec:knn_appendix}). We then split the re-sampled SEGUE-PS1 data randomly into test and training sets, where $70\%$ of the catalog was used to train the classifier and the other $30\%$ was used to test the classifier.  We performed ten iterations of training and testing, in the six-dimensional color space, and in each iteration we used the same classifier parameters but a different re-sampling of the SEGUE-PS1 data. We then averaged the output of the ten iterations, and re-normalized the individual label probabilities for each source so they summed to unity. All probabilities given hereafter are these averaged, re-normalized values. We adopted a $50\%$ probability threshold for classification, i.e. for a source to be assigned to a given label, the probability for that label must be higher than $50\%$.

The {\it completeness\/} of a sample resulting from a classifier is generally defined  to be the ratio of the number of sources that are correctly labeled to the true number of sources with that label (i.e., true positives divided by the sum of true positives and false negatives). Similarly, the {\it purity\/} is generally defined to be the ratio of the number of sources correctly labeled to the number of sources given that label (i.e., true positives divided by the sum of true positives and false positives). We found that, when applied to the test sets with known true stellar types, our  KNN classifier produces  samples of BHB that are, on average, $\sim 82\%$ complete and $\sim 81\%$ pure, together with samples of BSS that are, on average, $\sim 78\%$ complete and $\sim 80\%$ pure. These results are similar to, or better than, those obtained by previous  broad-band color based techniques reported in the literature, such as that of \citet{vickers} who used SDSS $griz$ photometry (they found $\sim 77 \%$ purity and $\sim 51\%$ completeness for a sample of BHB stars).

\begin{figure*}
\begin{center}
\includegraphics[width=.9\textwidth]{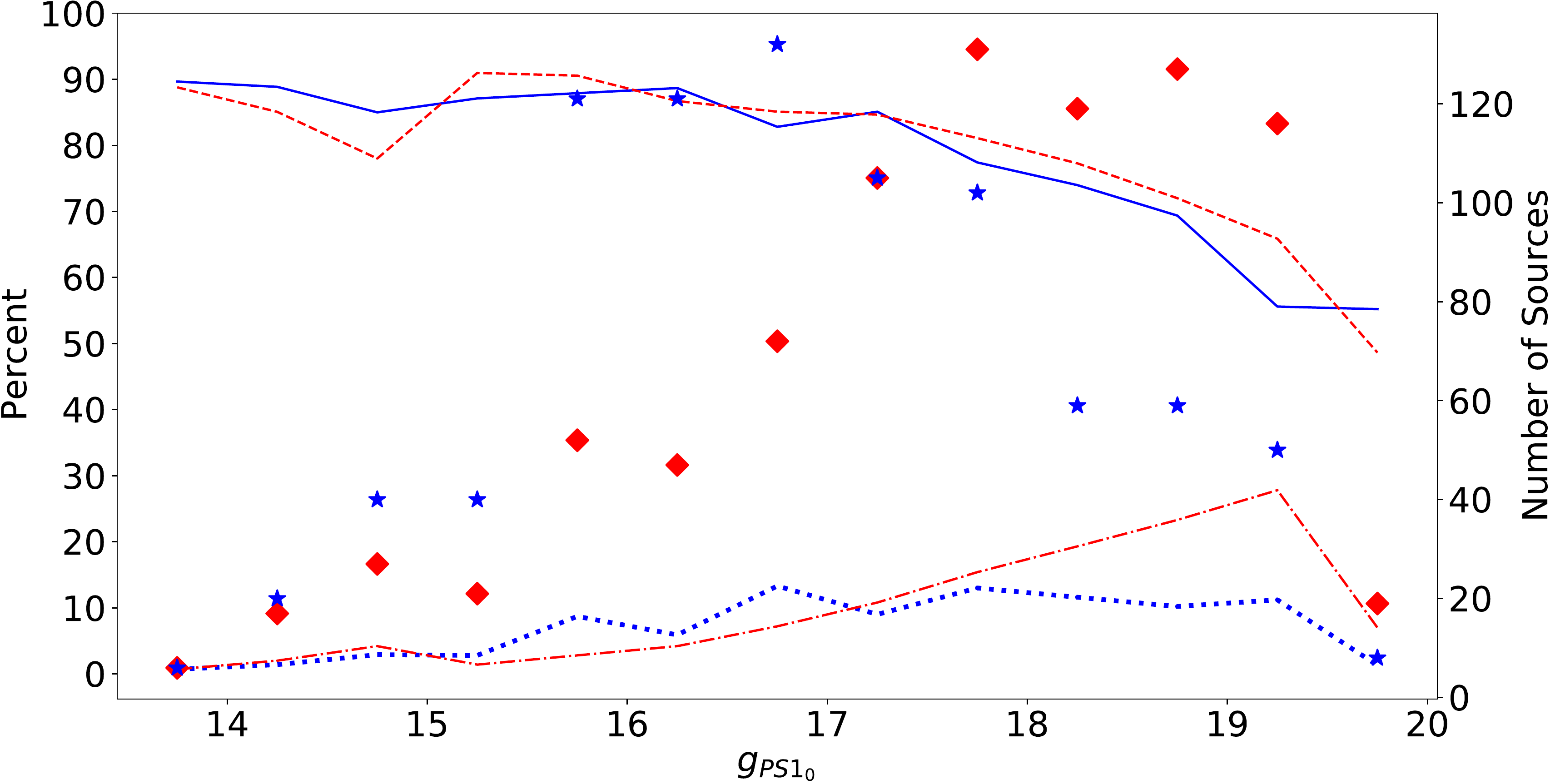}

\caption{The average percentage completeness of BHB (solid blue line) and BSS (dashed red line) retrieved in each 0.5 magnitude-wide bin of apparent magnitude, together with the average percentages of true BHB that get labeled as BSS (blue dotted line) and vice-versa (red dash-dotted line). Note that these percentages are not guaranteed to sum to 100\%, due to the small fractions of the BHB and BSS sources that were incorrectly classified as WD or QSO, or given no label with probability above 50$\%$. The blue stars and red diamonds represent the average number of BHB and BSS, respectively, in the test sets. In all cases, the values for bin centers are shown.}

\label{fig:knn_verification}
\end{center}
\end{figure*}

The KNN classifier uses only color information, but as photometric errors are larger  at fainter magnitudes, it is to be anticipated that its performance would decline with increasing apparent magnitude, as the larger uncertainties blur the distinction between BHB and BSS in color space. For reference, the average error in $(g_{PS1}-r_{PS1})_0$ color for blue sources at $g_{PS1_0} \sim 19$ is $0.02$~mag, increasing to  $0.04$~mag at $g_{PS1_0} \sim 20$ and to $0.06$~mag at $g_{PS1_0} \sim 21$. We found that performance did indeed decrease towards fainter magnitudes, as illustrated in Figure~\ref{fig:knn_verification}, which shows the percentage completeness  and false negatives for the selection of  BHB and BSS as a function of  apparent $g-$band magnitude.

The purity achieved  is of course sensitive to the relative numbers of BHB and BSS, as each is the main potential contaminant of the other (\citealp[see, for example Fig.~5 of][]{starkenburg}). The average number of true BHB and BSS in each half-magnitude bin of the test sets is indicated by the points (stars and diamonds, respectively) in Fig.~\ref{fig:knn_verification}.  These are determined by the structure of the Galaxy (a BHB star with an apparent $g-$magnitude of $19.75$ must be at a distance of $\sim 90 $ kpc, while a BSS of the same apparent magnitude could be much closer), modulated by the complex selection function that results from first, the SEGUE survey, second, the cross-match described above, and third, the re-sampling discussed above. The populations of BHB and BSS in \boo\ are most likely different from those in the training set, which has been adjusted to have equal proportions of BHB and BSS. \citet{momany} reports about twice as many BSS (of a broad range in magnitude) as (B)HB stars within the half-light radius  of  \boo,  using the photometry from \citet{belokurov06} and assigning a star to a given type based purely on its location in the observed CMD. The BHB and BSS selected this way are separated in apparent magnitude, so it is difficult to make a comparison with the results in  Fig.~\ref{fig:knn_verification}. Further,  contamination of the CMD by Milky Way stars is difficult to quantify. Happily, Milky Way BHB and BSS stars can largely be removed from our sample of \boo\ candidate members using astrometry.

\subsection{Additional Tests of the KNN Classifier}\label{sec:knn_comp}

We applied the KNN classifier to two datasets of known BHB stars (with spectra), the first being a different selection of BHB stars from SDSS/SEGUE by \citet{xue_2011} and the second being BHB with space motions consistent with membership of \boo, taken from \citet{koposov} and \citet{jenkins}.

\vfill
\eject

\subsubsection{The \citet{xue_2011} Sample}

\cite{xue_2011} identified candidate BHB stars in a cross-matched sample of SDSS photometry and spectroscopy, using cuts based on both colors and the shapes  of the  Balmer lines (H$\delta$ and H$\gamma$). We cross-matched their catalog of candidate BHBs to the PS1 photometry, following the same procedure as for our own  SEGUE-PS1 photometric  catalog, yielding a sample of  4538 candidate BHB stars with $-0.5 \le (g_{PS1}-r_{PS1})_0 \le 0$. Application of our KNN classifier to this sample resulted in  $89\%$ being labeled as BHB (adopting the $50\%$ probability threshold from above). Approximately $9\%$ of the stars in the sample are labeled as BSS stars, with the majority of the remaining stars ($\sim 2\%$ of the sample) being ambiguous, in that the classifier could not assign a probability higher than $50\%$ to any label.

Under the assumption that the \citet{xue_2011} catalog is entirely BHB stars (i.e.~$100\%$ pure), the completeness obtained using our KNN classifier is comparable with that we obtained earlier.  However, \citet{lancaster} identified likely BSS candidates within the  \cite{xue_2011} catalog, at the level of $\sim 10\%$ contamination and suggested refined  restrictions in color and in the  Balmer-line profiles to isolate the true BHB.  We found that $\sim 7\%$ of the cross-matched sample would be identified as BSS using the criteria proposed in \cite{lancaster} (ignoring their metallicity cut). Further, $48\%$ of the stars that our KNN classifier identified as BSS would be similarly classified by \citet{lancaster}, while 2$\%$ of those stars that our classifier identified as BHB would be identified as likely BSS with these refined criteria. This test revealed that some of the sources in the \citet{xue_2011} catalog that were ``mis-classified'' by our approach as BSS sources are, in fact, likely to be genuine BSS stars, and that the  sample of BHB stars identified by our classifier has only  a low level of contamination by these likely BSS stars, even though these BSS stars have properties similar enough to true BHB to be labeled as such using other techniques.

\subsubsection{Spectroscopic BHB members of \boo}

\begin{figure*}
\begin{center}
\includegraphics[width=.9\textwidth]{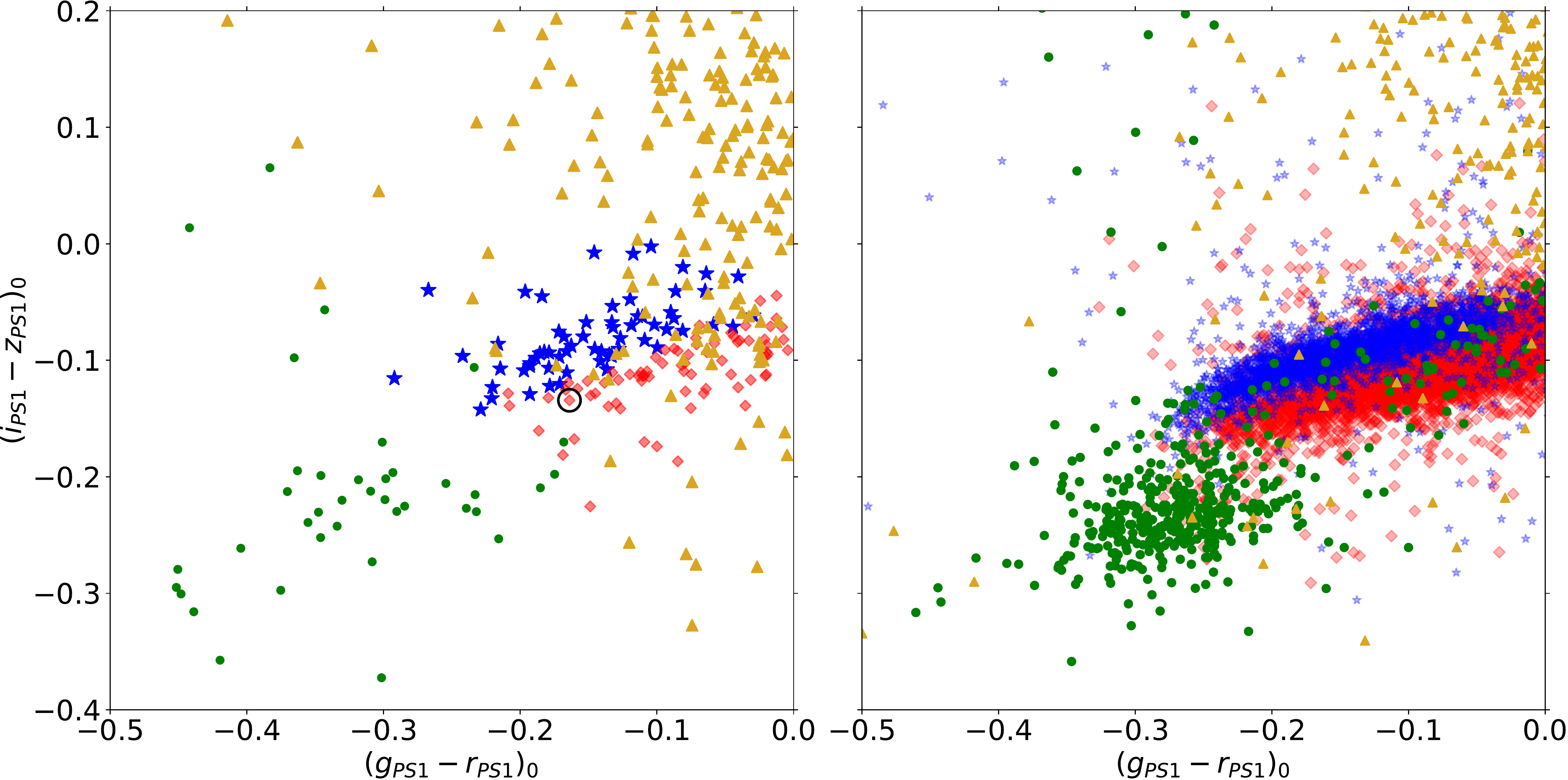}

\caption{The $(g_{PS1}-r_{PS1})_0, (i_{PS1}-z_{PS1})_0$ distribution of the labeled \boo\ catalog sources (lefthand panel), compared to the distribution of the SEGUE-PS1 cross-matched catalog sources (righthand panel). BHB are shown as blue stars, BSS/MS as red diamonds, WD as green circles, and QSOs as orange triangles. For clarity, the size and opacity of the symbols representing the BSS and BHB sources on the SEGUE-PS1 plot are reduced relative to those of the \boo\ field plot. The BHB from the \cite{koposov} and \cite{jenkins} sample that is mis-classified as a BSS star is indicated with an open black circle in the lefthand panel.}

\label{fig:boo_lab}
\end{center}
\end{figure*}

We cross-matched the catalog of sources observed spectroscopically by \citet{koposov} (re-analyzed by \citealp{jenkins}) with the PS1 photometric dataset, following the same criteria for matching radius and photometric quality   outlined above.  All six of the non-variable BHB candidates (based on location in the CMD) targeted by \citet{koposov} met the membership criteria of \citet{jenkins}, including having consistent proper motion from \textit{Gaia\/} EDR3, and all  six have matches with high-quality photometry in PS1. Application of the KNN classifier to these six sources resulted in  five of the stars being labeled as BHB and one being labeled as a BSS star. This $83\%$ completeness is consistent with the overall performance of the classifier, and is higher than the completeness seen for Milky Way halo BHB stars at a comparable faint magnitude, $g_{PS1_0} \sim 19.5$, in Figure~\ref{fig:knn_verification}. The star mislabeled as a BSS star lies in a region of color-color space occupied by both BSS and BHB (its location is marked by the open black circle symbol in the lefthand panel of Figure~\ref{fig:boo_lab}) and indeed, approximately ten times as many BSS than BHB in the training set have similar colors (defined as all six colors being within $\pm 0.015$ of those of the  misclassified BHB). 

Perhaps unsurprisingly given their photometric variability, the classifier also mislabeled the 5 RR Lyrae associated with \boo\ that had blue enough PS1 colors and sufficiently high-quality photometry to be included in the sample: 
three were classified as QSO, one as a BSS, and one was unclassified (no label was assigned with probability above 50\%).

\subsection{Selection of Blue Stars in a Wide-Area  Bo{\"o}tes I Field}

We investigated the existence of an extended stellar envelope of  \boo\ by first  selecting all  sources with high-quality 4-band  PS1 photometry,  and with positions within a four-by-four degree box around the coordinates of the nominal center of \boo. The area selected encompasses more than $\sim 100$ times that enclosed by the (elliptical) half-light isophote of \boo\ ($\simlt 0.1$~square degree, based on the exponential fit to the surface density profile from \citealt{munoz}) and allows exploration of the stellar content well  beyond the King profile nominal tidal radius of \boo.   We imposed the same quality-control requirements on the photometry as described above, ensuring that each source has point-like photometry in each filter, and again corrected the photometric data for dust reddening and extinction. The resultant photometric dataset is composed of sources brighter than $g \sim 22$ ($\sim 2.5$ magnitudes fainter than the expected magnitude of BHB  at the distance of Boo I, see below) and is hereafter referred to as \lq the \boo\ catalog’. Any possible non-uniformity in the limiting magnitude of this catalog should affect the completeness of only the BSS sample, and these sources are sufficiently faint that the decreasing performance of the KNN classifier with apparent magnitude significantly impacts the completeness  (see Figure \ref{fig:knn_verification}). Happily, our science goals do not require a complete sample of BSS and we did not attempt to correct for incompleteness.

We then applied the (trained) KNN classifier to the \boo\ catalog,  again requiring a minimum probability of $50\%$ for a given label to be assigned to a source. Of the 437 sources in the \boo\ catalog with $-0.5 \le (g_{PS1}-r_{PS1})_0 \le 0$, the classifier  labeled 62 as BHB and 76  as BSS (see Table~\ref{tab:of-param}). Of the remaining sources, 52 were labeled as WD, 225 were labeled as QSO, and 22 were ambiguous.  Fainter than $g_{PS1_0} \sim 21$, no  source was labeled as either BHB or BSS. This likely reflects the trend of decreasing completeness at fainter magnitudes seen in Figure~\ref{fig:knn_verification}.

The distribution of all the classified star-like sources in the $(g_{PS1}-r_{PS1})_0, (i_{PS1}-z_{PS1})_0$ color-color plane is shown in Figure~\ref{fig:boo_lab}, where the lefthand panel displays the sources in the \boo\ catalog that were successfully labeled (i.e.~with probability greater than 50\%) and the righthand panel displays the distribution of the  SEGUE-PS1 cross-matched catalog, in which case the labels were based on the SEGUE spectroscopy, as described above in Section~\ref{sec:photometry_selection}.  The loci occupied by the different types of source are similar in the two panels, albeit that there is a higher relative number of QSO in the lefthand panel (the \boo\ catalog). This increased presence of QSOs plausibly reflects the combined effects of the higher fraction of extragalactic sources compared to Galactic stars at the deeper limiting magnitude of the \boo\ catalog, the increased misclassificaton of sources at fainter magnitudes due to larger photometric errors, and  the SEGUE selection function. The majority ($68\%$) of these sources labeled as QSO are fainter than $g_{PS1_0} = 20$, the limiting magnitude of the SEGUE-PS1 catalog.

The subset of sources classified as either BHB or BSS in the \boo\ catalog includes field  stars in the Milky Way, in addition to actual members  of \boo.  The Milky Way stars should span a broad range of apparent magnitudes, reflecting their (unknown) distances, while the member stars of \boo\ should be constrained to a narrower range of apparent magnitudes, consistent with the known distance to \boo\ and their intrinsic luminosities.

The large metallicity spread of \boo\ ($\Delta [\rm{Fe}/\rm{H}] \sim 1.7$~dex, e.g. \citealt{norris}) should cause some spread in the absolute magnitude of BHB (at fixed color). The recent calibration by \citet[][their equation 5]{fermani}, valid over the range $-3 \simlt  {\rm [Fe/H]} \simlt -1$, would imply a spread of $\sim 0.15$~mag in absolute $g$-magnitude at a color of $(g-r)_{SDSS} = -0.15$, for a metallicity distribution similar to that of \boo. Empirically, the BHB fiducials of globular clusters of very similar iron abundances vary by  a similar, or even larger amplitude in absolute magnitude - for example, \citet{bernard} found a range of $\simlt 0.3$~magnitudes for the BHB in a set of four metal-poor globular clusters, with iron abundance varying by only $\sim 0.1$~dex. Further, the uncertainty in the distance estimate given by \citet{okamoto} ($65 \pm 3$~kpc), translates to an uncertainty of 0.2 in apparent magnitude, for given absolute magnitude. Given these considerations, we decided to define the apparent magnitude range of the BHB in \boo\ by the 13~Gyr, ${\rm [M/H] = -2.2}$ PARSEC isochrone, shifted by a distance modulus corresponding to 65~kpc, and allowing a spread of $\pm 0.2$ magnitudes, vertically, about the locus of the HB set by the isochrone (the uncertainty in the PS1 photometry at the magnitudes of interest, indicated on Figure~\ref{fig:panstarrs_cmd}, is small in comparison). We confirmed that all of the six BHB members from \citet{jenkins} lie within these boundaries.

The BSS systems of a given stellar population will have a range of absolute magnitudes, depending on the (uncertain) details of their formation mechanism(s), but for an old population \citealt{hurley,geller} they should be  fainter than the horizontal branch and indeed be within a couple of magnitudes brighter than the main-sequence turnoff (MSTO). This expectation is consistent with the predicted brightness of an unevolved main-sequence star of mass equal to twice that of the old turn-off, which is $\sim 2$~magnitudes brighter than the turn-off in the $g-$band ($1.93$ magnitudes with our fiducial isochrone). Further, the observed BSS populations of globular clusters, in the cases where proper motions are available to confirm membership, are within $\sim 2.5$~magnitudes of the turn-off \citep{baldwin}. The MSTO of \boo, defined by the bluest point of our fiducial isochrone, corresponds to  $g_{PS1 _0} = 23.07$.  Adopting the metal-poor globular cluster M92 as an empirical fiducial and using the data from \citet{bernard} predicts a MSTO for \boo\ of $g_{PS1 _0} = 23.1$, in excellent agreement. We therefore identified candidate BSS members of \boo\ by requiring that they have apparent magnitude fainter than $g_{PS1 _0} = 23.1 - 2.5 = 20.6$ (this cut removes the mis-classified BHB and RR Lyrae members). Note that the faintest BSS in our sample should then be at the   $g_{PS1 _0} \sim 22$ limit set by the stringent quality controls we implemented on the multi-band photometry.

\begin{figure*}
\begin{center}
\includegraphics[width=.9\textwidth]{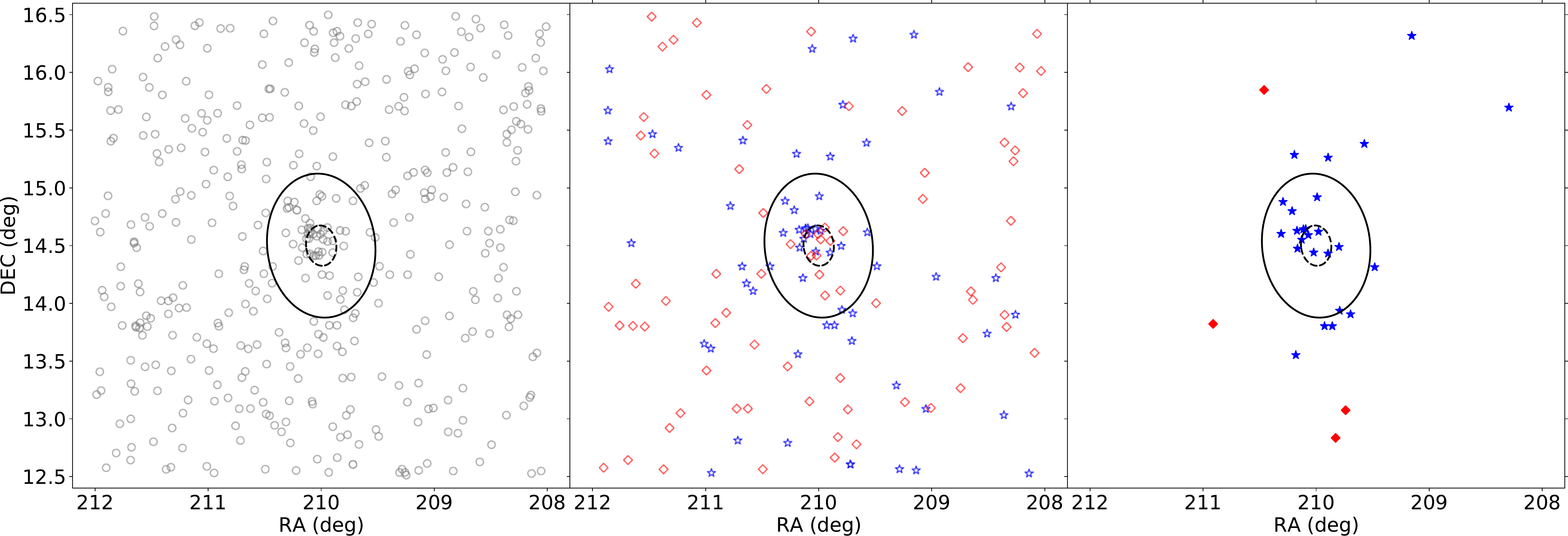}

\caption{The two-dimensional spatial distribution of various subsets of blue sources in the field of view of \boo, together with the King profile tidal and exponential half-light ellipses of \cite{munoz} (black solid and dashed lines, respectively). The leftmost plot shows all sources with $-0.5 \le (g_{PS1}-r_{PS1})_0 \le 0$, the middle panel shows all BHB or BSS stars, and the rightmost panel shows the subset of BHB and BSS stars that have apparent magnitudes that are consistent with the expected magnitudes of BHB and BSS at the distance of \boo. }

\label{fig:blue_dist}
\end{center}
\end{figure*}

\begin{figure}
\begin{center}
\includegraphics[width=.45\textwidth]{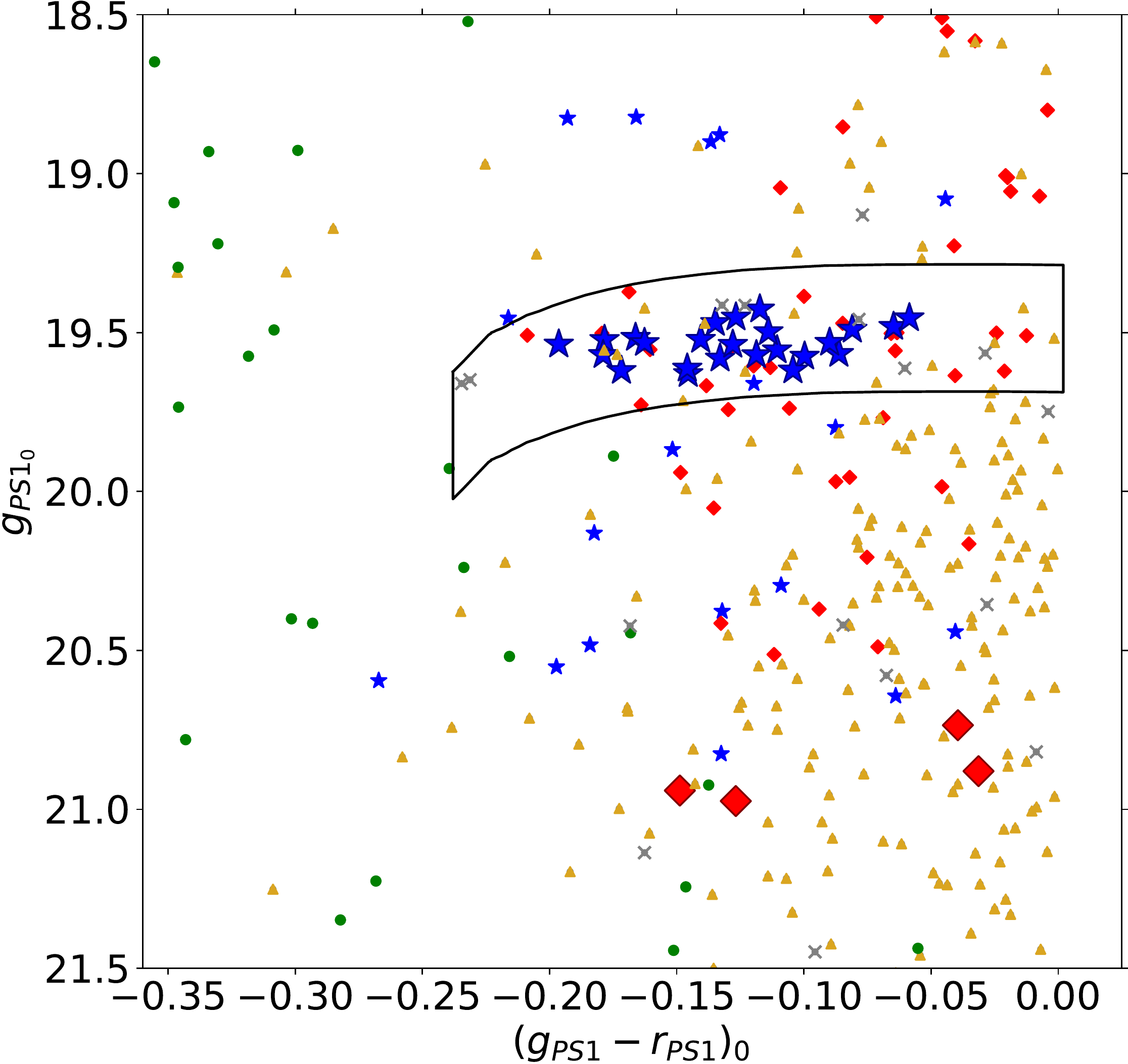}

\caption{The CMD of blue sources in the \boo\ field, color-coded by the label assigned by the classifier. BHB stars are shown as blue stars, BSS as red diamonds, WD as green circles, QSO as orange triangles, and ambiguous sources with no label having higher than $50\%$ probability are shown as grey crosses. The isochrone-based selection region used to identify BHB stars with appropriate apparent magnitudes is outlined in black. We verified via proper motion that the bright BHB just above this region is unlikely to be a member of \boo. The BHB that do have consistent proper motions are shown with larger symbols, as are the BSS that pass our imposed magnitude cut for potential members ($g_{PS1_0} > 20.6$). The sources with ambiguous classifications within the BHB selection region are discussed further in the text.}

\label{fig:labeled_cmd}

\end{center}
\end{figure}

There are 25 candidate BHB and 4 candidate BSS selected by this step, with the projected spatial distribution as shown  in Figure~\ref{fig:blue_dist} (righthand panel). The other panels of Fig.~\ref{fig:blue_dist} show, for comparison, the distribution of all the blue sources in the \boo\ catalog (lefthand panel) and that of all sources, independent of apparent magnitude, labeled as either BHB or BSS (middle panel). The selection region used to delineate the appropriate ranges of apparent magnitudes and colors for BHB stars at the distance of \boo\ is shown in Figure~\ref{fig:labeled_cmd}, over-plotted on the CMD of the blue sources in the \boo\ field. Each different class of source, as labeled by our KNN classifier, is indicated by symbol type and color (as described in the figure caption). The final set of  candidate BHB member stars (including proper-motion cuts as described below) and those apparent magnitude-appropriate BSS are indicated with larger symbols. 

Some  Milky Way field stars could make it through these photometric cuts and we implemented further astrometric cuts, as described in  Section~\ref{sec:gaia_selection}.

\subsection{Selection of Blue Stars in Comparison Off-Fields} \label{sec:off_field_phot}
\begin{figure}

\begin{center}

\includegraphics[width=.45\textwidth]{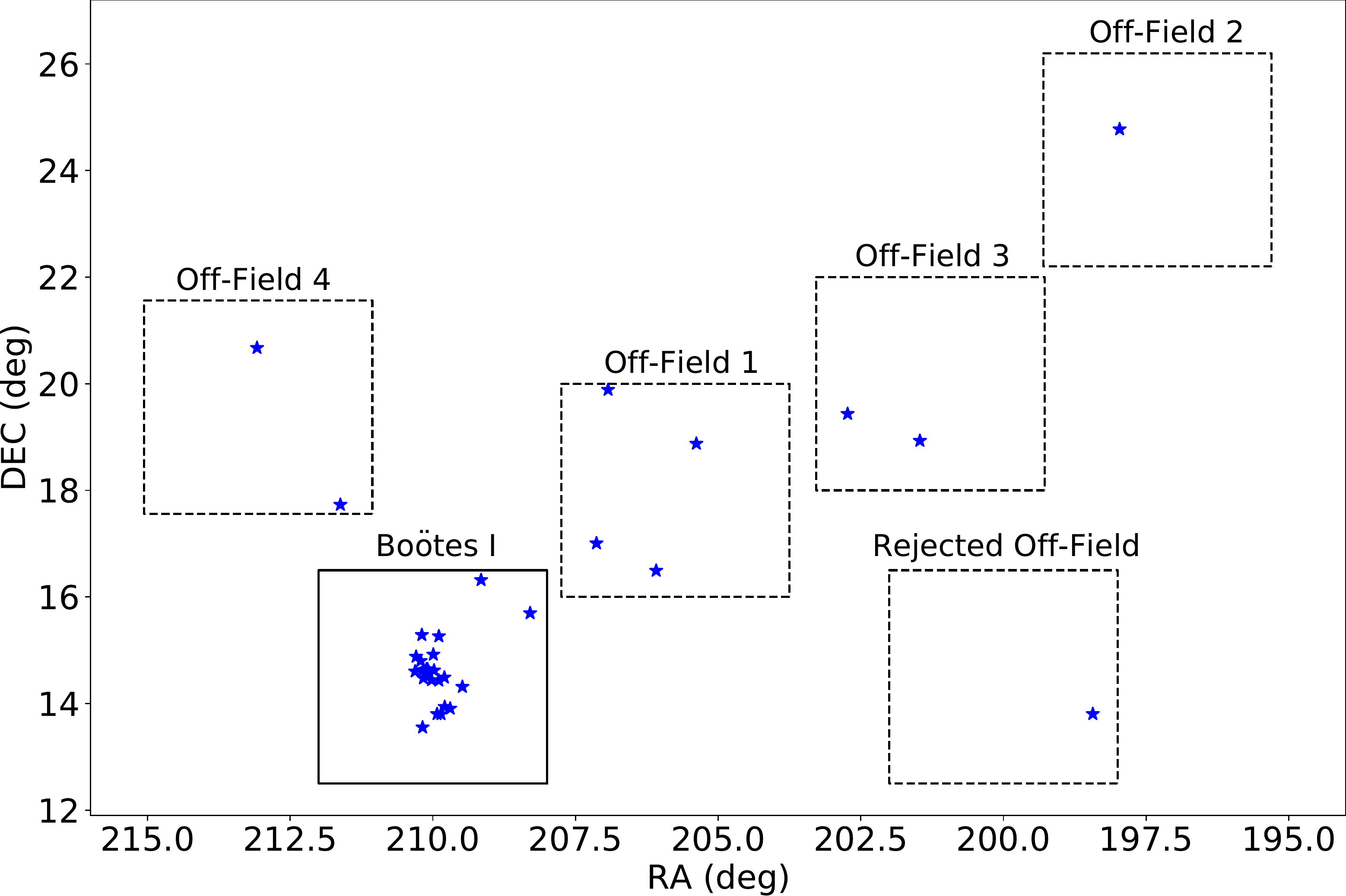}

\caption{The on-sky footprints of the four main off-fields considered in this work, together with the \boo\ field and the additional rejected off-field that is contaminated by stars from the Sagittarius stream. The BHB stars that have both proper motions and apparent magnitudes consistent with membership of \boo\ are shown as blue stars.}

\label{fig:field_footprints}

\end{center}
\end{figure}

We selected four lines-of-sight at comparable Galactic coordinates to \boo, to define \lq off fields' within which to estimate the contamination of the sample of BHB and BSS obtained above, by non-members of \boo. \boo\ is located in the \lq Field of Streams' \citep{FoS} and we took considerable care to minimize  contributions in the off-fields from the Sagittarius (Sgr) streams, by inspection of the proper motions of all sources in the field (a nearby fifth field which shows a clear signature of Sgr in the proper motions is discussed in Appendix~\ref{sec:Sgr}).  The coordinates of the centers of the off-fields we selected are given in Table~\ref{tab:of-param} and the locations on the sky of these $4 \times 4$~square~degree fields, plus the fifth  (Sgr stream) field, are illustrated in Figure~\ref{fig:field_footprints}. From analytic models of Milky Way stellar density profiles, we would not expect significant dependence of the surface density of faint blue stars on Galactic latitude over the range of field-center coordinates considered in this work. Indeed, the TRILEGAL models \citep{trilegal} predict only a $\lesssim 20\%$ difference of the surface densities of faint blue stars (which we defined to be $-0.5 \le (g_{PS1} - r_{PS1}) \le 0$, $19 < g_{PS1} < 21$ in the extinction-free, error-free synthetic magnitudes provided by TRILEGAL) between fields centered on $(\ell, b) = (355, 70)$ and $(\ell, b) = (355, 85)$. We thus proceeded with the selected observational off-fields.

We applied the KNN classifier to the blue star-like sources in each off-field, after obtaining photometric catalogs following the  procedure detailed  above. The resultant numbers\footnote{ As a consistency check, we note that the number of sources labeled as QSO in each of the off-fields is $\sim 200$, approximately the same as for the \boo\ field, and again as in the \boo\ field, $\sim 70\%$ of the sources labeled as QSO are fainter than $g_{PS1_0}=20$.} of BHB and BSS are given in Table~\ref{tab:of-param}, where the two rightmost columns indicate the numbers of BHB and BSS with appropriate apparent magnitudes to be at the heliocentric distance  of \boo\ (assumed to be 65~kpc; \citealp{okamoto}). Inspection of Table~\ref{tab:of-param} shows that the number of stars in each off-field that are classified as BHB and have photometry consistent with being at the distance of \boo\ is approximately constant field-to-field (with the exception of off-field~1), at a level of $10 - 20\%$ of the BHB in the \boo\ field. The implied surface density of BHB in the field stellar halo of $\sim 1$ BHB per 4~square degree at a distance of $\sim 65$~kpc is somewhat higher than the count of 3 BHB in 40 square degree, at distances of $50-100$~kpc, found by \citet{deason} in lines-of-sight away from both the Sgr stream and the VVDS over-density. This apparent disagreement is perhaps unsurprising, given that the selected off-fields are in close proximity to known over-densities, which were explicitly avoided in \cite{deason}. It may be expected that such field halo BHB should be on different orbits about the Galactic center than are  the BHB members of \boo, even if at similar Galactocentric distance, and so should be distinguishable by kinematics, which we turn to next.

\begin{deluxetable*}{cccccccccc}
\tablecolumns{11}
\tablewidth{0pt}
\tablecaption{Field Centers and Blue Star Counts \label{tab:of-param}}
\tablehead{
\colhead{Name} &  \colhead{$\ell$ }&  \colhead{$b$} &\colhead{RA} & \colhead{Dec} & \colhead{All Blue} & \colhead{All BHB} & \colhead{All BSS} & \colhead{BHB} & \colhead{BSS} \\
\colhead{} &  \colhead{} &  \colhead{} &\colhead{(J2000)} & \colhead{(J2000)} & \colhead{} & \colhead{} & \colhead{} & \colhead{} & \colhead{}
}

\startdata
\boo & 358.0$^\circ$  & 69.6$^\circ$  & 210.00$^\circ$  & 14.50$^\circ$  & 437 & 62 & 76 & 25 & 4 \\
Off-Field 1 & 358.1$^\circ$  & 75.0$^\circ$  & 205.75$^\circ$ & 18.00$^\circ$  & 389 & 42 & 53 & 6 & 2\\
Off-Field 2 & 357.7$^\circ$ & 85.0$^\circ$  & 197.30$^\circ$  & 24.20$^\circ$ & 331 & 48 & 37  & 2 & 1\\
Off-Field 3 & 352.0$^\circ$ & 79.5$^\circ$  & 201.28$^\circ$  & 20.00$^\circ$ & 424 & 47 & 55  & 4 & 1\\
Off-Field 4 & 15.0$^\circ$ & 70.0$^\circ$ & 213.06$^\circ$  & 19.56$^\circ$ & 344 & 42 & 52 & 2 & 1\\
\enddata

\tablecomments{All fields are $4 \times 4$~sq.~degree. The column \lq All Blue' gives  the number of star-like sources within that field with high-quality PS1 photometry and with $-0.5 \le (g_{PS1}-r_{PS1})_0 \le 0$. \lq All BHB' and \lq All BSS' give the subsets of the blue stars that are so classified, and \lq BHB' and \lq BSS' give the further subsets that have  appropriate apparent magnitudes to be at the distance of \boo. 
}
\end{deluxetable*}

\section{Astrometric Identification of Likely Boo~I Members}\label{sec:gaia_selection}
The \textit{Gaia} EDR3 astrometric catalog is sufficiently precise and accurate  at the apparent magnitudes of interest that it can be utilized both to determine the mean proper motion of \boo\ and then to isolate potential Milky Way contaminants in the photometrically selected, apparent-magnitude consistent BHB sample. We first cross-matched the PS1 photometric catalogs for each of the fields in Table~\ref{tab:of-param} with  the \textit{Gaia} EDR3 data (querying the \texttt{gaia\_source} catalog), using the functionalities provided in \texttt{TOPCAT} \citep{topcat}. We required that the \textit{Gaia} data for each source were well-fit by a single-star model (\texttt{ruwe} $< 1.3$), the astrometric solution was derived over at least 6 visibility periods (\texttt{visibility\_periods\_used} $> 6$), and the photometry was not significantly affected by  background light or blending (thus requiring that the photometry satisfy the relation $1.0 + 0.015(\rm{G_{BP} } - \rm{G_{RP}})^2 < $ \texttt{phot\_bp\_rp\_excess\_factor }$< 1.3 + 0.06(\rm{G_{BP} } - \rm{G_{RP}})^2$, as recommended by\footnote{This requirement is also equation C.2. of \cite{lindegren2018}.} \citealp{gaia-HRD}).

\subsection{Derivation of the Mean Proper Motion of Bo{\"o}tes I}\label{sec:mean-pm} 

We derived our own estimate of the mean proper motion of \boo, in order to  create an internally consistent sample of proper-motion members, from which we defined a covariance ellipse that we used to identify new, astrometrically consistent, candidate members. We defined the mean proper motion using the \textit{Gaia} EDR3 astrometry (\citealt{gaia_2016}, \citealt{gaia_2020}) for stars that had been previously identified, primarily through spectroscopy,  as members.  We used the samples of \cite{martinspec}, \cite{norris}, and \cite{koposov} to compile a catalog of stars that have, at a minimum, line-of-sight velocity consistent with membership. We supplemented this with the full set of 16  RR Lyrae variables identified by \cite{siegel} and/or \cite{vivas} as being associated with \boo; any found not to have consistent proper motions would be removed though the process described below (as noted above the \textit{Gaia} DR2 proper motions suggested one was a non-member). Most of this combined sample lies within the half-light radius of \boo, although two members on the RGB from \citet{norris} and two of the RR Lyrae stars lie beyond the King profile nominal tidal radius.

There is some overlap among the  spectroscopic  samples and there is not always agreement in the published results as to the membership status of a given star. We proceeded by assigning our own membership value of unity to every star indicated as a member of \boo\ in each study, and a membership value of zero to stars indicated as non-members in that study. The spectroscopic data for the sample observed by  \cite{koposov} was re-analyzed by \citet{jenkins} and  we adopted the membership criteria developed by the latter authors.   We averaged the  membership values for any star in more than one sample, removed the duplicate entries  and matched the resulting catalog of unique stars to both the PS1 photometry and the \textit{Gaia} EDR3 catalog.

We  considered all stars with an averaged  membership value of $\ge 0.5$ to be actual members of \boo\  and  we  derived the mean proper motion of \boo\ from only the subset of these members brighter than $G_{Gaia} = 20$, for which the  proper motion uncertainties are $\simlt 0.5$~mas~yr$^{-1}$.  We note that none of these member stars has a robust parallax measurements in \textit{Gaia} EDR3 (all have have parallax errors larger than $20\%$), indicating that none is likely to be a nearby foreground star. We calculated the median of the proper motions in each of right ascension and declination ($\mu_{\alpha *}, \, \mu_{\delta}$, with the usual convention that the $*$ indicates the cosine of declination), and then removed  all stars with proper motions outside of the three-sigma covariance ellipse defined by the initial sample of bright members.  We then repeated this process, iteratively defining the median and rejecting stars outside of the new three-sigma covariance ellipse, until no further stars were rejected. We used the final set of surviving members (59 stars) to determine the mean proper motion of \boo, which we defined as the error-weighted average of the proper motions of each of those stars. We further  adopted the three-sigma covariance ellipse from this final set of stars as the proper motion ellipse of \boo.

The mean proper motion of \boo\ defined by this iterative  process is $\mu_{\alpha *} = -0.397 \pm 0.019$~mas~yr$^{-1}$, $\mu_{\delta} = -1.066  \pm 0.015$~mas~yr$^{-1}$. These values are  in excellent agreement with the \textit{Gaia} EDR3-based proper motion determination of \cite{mcconnachie_edr3} ($\mu_{\alpha *} =  -0.39 \pm 0.01$~mas~yr$^{-1}$, $\mu_{\delta}= -1.06 \pm 0.01$~mas~yr$^{-1}$) and is within two sigma of the EDR3-based proper-motion values found by \cite{li} ($\mu_{\alpha *} = -0.307 \pm 0.052$~mas~yr$^{-1}$, $\mu_{\delta} =-1.157 \pm 0.043$~mas~yr$^{-1}$),  by \cite{martinez_garcia} ($\mu_{\alpha *} = -0.357 \pm 0.029$~mas~yr$^{-1}$, $\mu_{\delta} = -1.071 \pm 0.024$~mas~yr$^{-1}$), and by \cite{battaglia} ($\mu_{\alpha *} =  -0.369 \pm 0.039$~mas~yr$^{-1}$, $\mu_{\delta}= -1.071 \pm 0.029$~mas~yr$^{-1}$). We used the mean proper motion (plus line-of-sight velocity and distance) to derive orbital parameters for \boo\ in two different Milky Way potentials, as described in  Appendix \ref{sec:dynamical}. It should be noted that the favored  orbit for \boo\ adopted   in the analysis of \citet{fellhauer} is remarkably similar to that found here, despite no proper-motion information being available to those authors.

\subsection{Selection of Proper-Motion Consistent Member Stars}


We defined a star to be ``proper motion consistent'' provided that the error ellipse of its reported proper motion  intersected with, or was contained by,  the three-sigma covariance ellipse defined by the set of 59 stars used above to determine the mean proper motion of \boo.  We found that approximately $30\%$ of all the blue stars, independent of apparent magnitude range,  in the \textit{Gaia}-PS1 cross-matched catalog for the \boo\ field have proper motions consistent with membership of \boo, compared to only $\sim 20\%$ of all the blue stars in each of the \textit{Gaia}-PS1 catalogs for the off-fields. As such, the majority of non-member contaminants can be identified by their proper motions ($\sim 80\%$, assuming all stars in the off-fields are non-members), which allows for the refinement of the candidate-member catalog that was obtained based on  photometry alone.

\subsubsection{BHB Candidate members in the Boo~I Field}

Of the 25 stars in the \boo\ field that were identified by the KNN classifier as candidate BHB and which also have  appropriate magnitudes to be at the distance of \boo,  24 have consistent proper motions. We verified that none of these 24 stars has a robust measurement of parallax in \textit{Gaia} EDR3 all have parallax error larger than $20\%$) and thus none is likely to be a nearby foreground star. The BHB that was rejected on the grounds of having inconsistent proper motion is only slightly outside of our (rather generous) proper-motion bounds, and could be considered proper-motion consistent in the future, depending on improved astrometry from future \textit{Gaia} data releases. As seen in Figure \ref{fig:labeled_cmd}, there are seven ambiguous sources (indicated by grey crosses) with colors and magnitudes that place them within the selection region for BHB stars in \boo. Three of these ambiguous sources have proper motions consistent with \boo, two of which are located interior to the King profile nominal tidal radius, each with BHB label probability below $\sim 45\%$. The third source lies beyond the King profile nominal tidal radius but has only a $\sim 30\%$ probability of being a BHB star, and thus is unlikely to be a genuine BHB member of \boo.

Five of the six BHB stars from the \cite{koposov} and \cite{jenkins} sample survive the iterative clipping we used to define the three-sigma covariance ellipse, while  all six meet the criterion to be considered proper-motion consistent. This apparent inconsistency arises due to the fact that while the errors in proper motion are not considered in the definition of the three-sigma covariance  ellipse,  they are taken into account when considering whether or not an individual star could be a member. Thus five of the BHB stars have proper motion measurements that lie inside of the ellipse, while the proper-motion measurement of the sixth lies outside, but with an error ellipse that intersects the three-sigma covariance ellipse.

\subsubsection{Boo~I Field BSS Candidate members}

Our requirement that BSS candidates be no more than 2.5~magnitudes brighter than the dominant old MSTO, which for \boo\ corresponds to $g_{PS1_0} > 20.6$, means that they are unlikely to have \textit{Gaia} EDR3 astrometric data (limiting apparent magnitude $G_{Gaia} \sim 21$) of high enough quality for membership tests. This is indeed the case, and none of the candidate BSS members in the \boo\ field, identified using photometry alone,  can be included in the sample of proper-motion consistent candidate member stars.

\subsubsection{Variable Stars in Bo{\"o}tes I}\label{sec:LPVsec}

\begin{figure}

\begin{center}

\includegraphics[width=.475\textwidth]{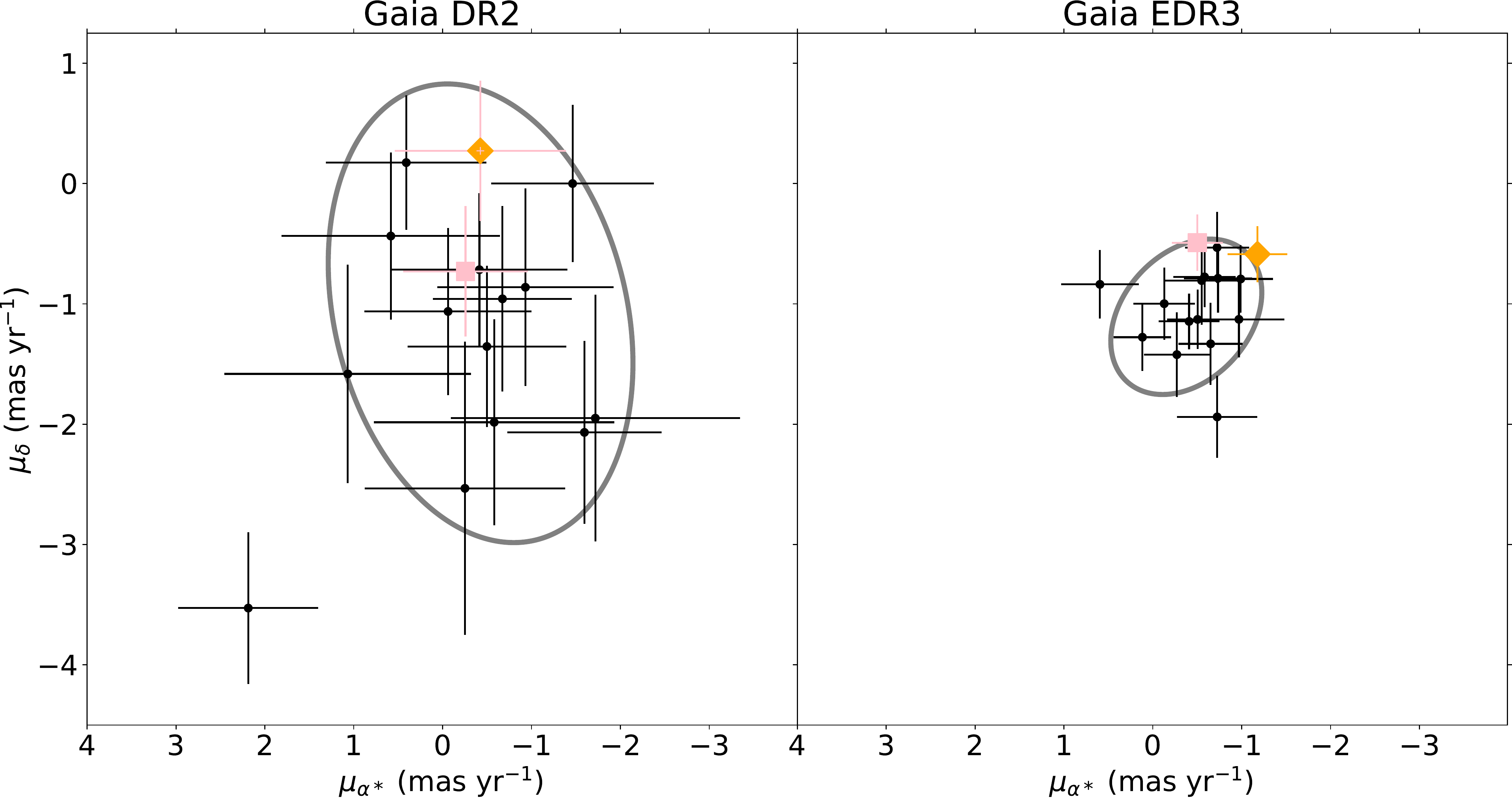}

\caption{The proper motions from \textit{Gaia} DR2 (lefthand panel) and EDR3 (righthand panel) for the RR Lyrae stars identified by \citet{siegel} and  \citet{vivas}. The error bars correspond to the quoted uncertainties in the proper motions for each source in the DR2 or EDR3 catalog. The three-sigma covariance ellipse for each dataset, defined through iterative sigma-clipping of previously identified member stars,  is shown in grey. The pink square indicates the outer envelope RR Lyrae star from \citet{siegel} while the orange diamond indicates the additional outer envelope RR Lyrae from \citet{vivas}.}

\label{fig:rrlyrae_comp}

\end{center}

\end{figure}

The 16 RR Lyrae variables associated with Boo~I by \cite{siegel} and by \cite{vivas} were included  in the initial stage of the iterative procedure we used to define the mean proper motion of \boo, and three of these stars were rejected in the sigma-clipping step. However, all 16 RR Lyrae stars have proper motions consistent with membership of \boo\ when their errors are considered (similar to the situation for the six BHB from \citealp{koposov} discussed above). This set of stars provides a graphic illustration of the improvement in the astrometric data for evolved stars in \boo\ between \textit{Gaia} DR2 and EDR3. Figure~\ref{fig:rrlyrae_comp} shows the comparison between the proper motions for the RR Lyrae based on DR2 (lefthand panel) and on EDR3 (righthand panel). The three-sigma covariance ellipse indicated on each panel was defined through the iterative sigma-clipping procedure described above and based on an input list of candidate members that included the RR Lyrae themselves. The EDR3 data yield reduced  scatter in the distribution of proper motions and considerably smaller  individual error bars. Further, the RR Lyrae star that was rejected by \citet{vivas}, based on DR2, is retained using EDR3. The two RR Lyrae at projected radii beyond the King profile nominal tidal radius are indicated by special symbols in Figure~\ref{fig:rrlyrae_comp} (a pink square for the candidate  from  \citealp{siegel};  an orange diamond for that from \citealp{vivas}). 

In addition to RR Lyrae variables, a candidate LPV member was identified by  \cite{siegel} and \cite{dallora}, as noted above. The coordinates of this LPV star place it within the half-light ellipse of \boo\ and we found its proper motion in  \textit{Gaia} EDR3 to be consistent with that of \boo, again indicating likely membership. We identified an object with matching coordinates in the general LAMOST catalog \citep[see][]{lamost1,lamost2}. The signal-to-noise of the spectrum (\texttt{obsid} 230716130) is sufficiently low that only a redshift is reported, with value   
$0.00028465 \pm 0.0000433$. The corresponding heliocentric velocity, $\sim 85 \pm 13$ km s$^{-1}$, is consistent with membership of \boo\ (mean velocity of $\sim 100$~km s$^{-1}$; \citealp{jenkins}), but this interpretation must be treated with caution due to both the low signal-to-noise ratio and the inherent difficulties/uncertainties in measuring the center-of-mass velocity for pulsating stars \citep[see, e.g.][]{lpv_rv}. We identified a source with matching coordinates in the 2MASS catalog (\texttt{2MASS} $13594873+1434483$). This source has a J-K$_s$ color of $0.8$, indicating that it is unlikely to be  an AGB star with a very dusty envelope (see e.g. \citealt{NW2000}). The most metal-poor LPV stars discovered to date reside within the globular cluster M15 \citep[NGC~7078;][]{mcdonald}, which has an iron abundance of  $[\rm{Fe/H}] = -2.29$ \citep{bailin}, some $\sim 0.3$ dex higher than the mean iron abundance of \boo\ (mean $[\rm{Fe/H}] = -2.6$, \citealp{frebel}). Should this LPV star be a true member of \boo, it could also be the most metal-poor LPV discovered. This exciting possibility warrants additional investigation.

\subsubsection{Bo{\"o}tes II}

The Bo{\"o}tes~II UFD (hereafter Boo~II) has central coordinates of  ${\rm (RA, Dec)  =  (209^\circ.50, 12^\circ.85)}$ (\citealt{walsh07}) and thus lies within the four-by-four degree \boo\ field explored in this work. Boo~II is closer than \boo, with a heliocentric distance of $42 \pm 1.6$~kpc \citep{walsh08}, and has a higher proper motion ($\mu_{\alpha*}, \mu_{\delta} = -2.273 \pm 0.151, -0.361 \pm 0.115$~mas~yr$^{-1}$, \citealt{li}). Contaminants from Boo~II within the sample of candidate \boo\ members is thus unlikely. Indeed, inspection of Figure~\ref{fig:blue_dist} shows that there are no apparent-magnitude consistent BSS or BHB within $\sim 0.3$~degree of the center of Boo~II ($\sim 6$~half-light radii, adopting the exponential half-light radius of $3.07 \pm 0.44$~arcmin for Boo~II from \citealt{munoz}).

\subsubsection{ Proper-Motion Consistent Blue Member Stars in the Boo I Field}

\begin{figure}
\begin{center}
\includegraphics[width=.45 \textwidth]{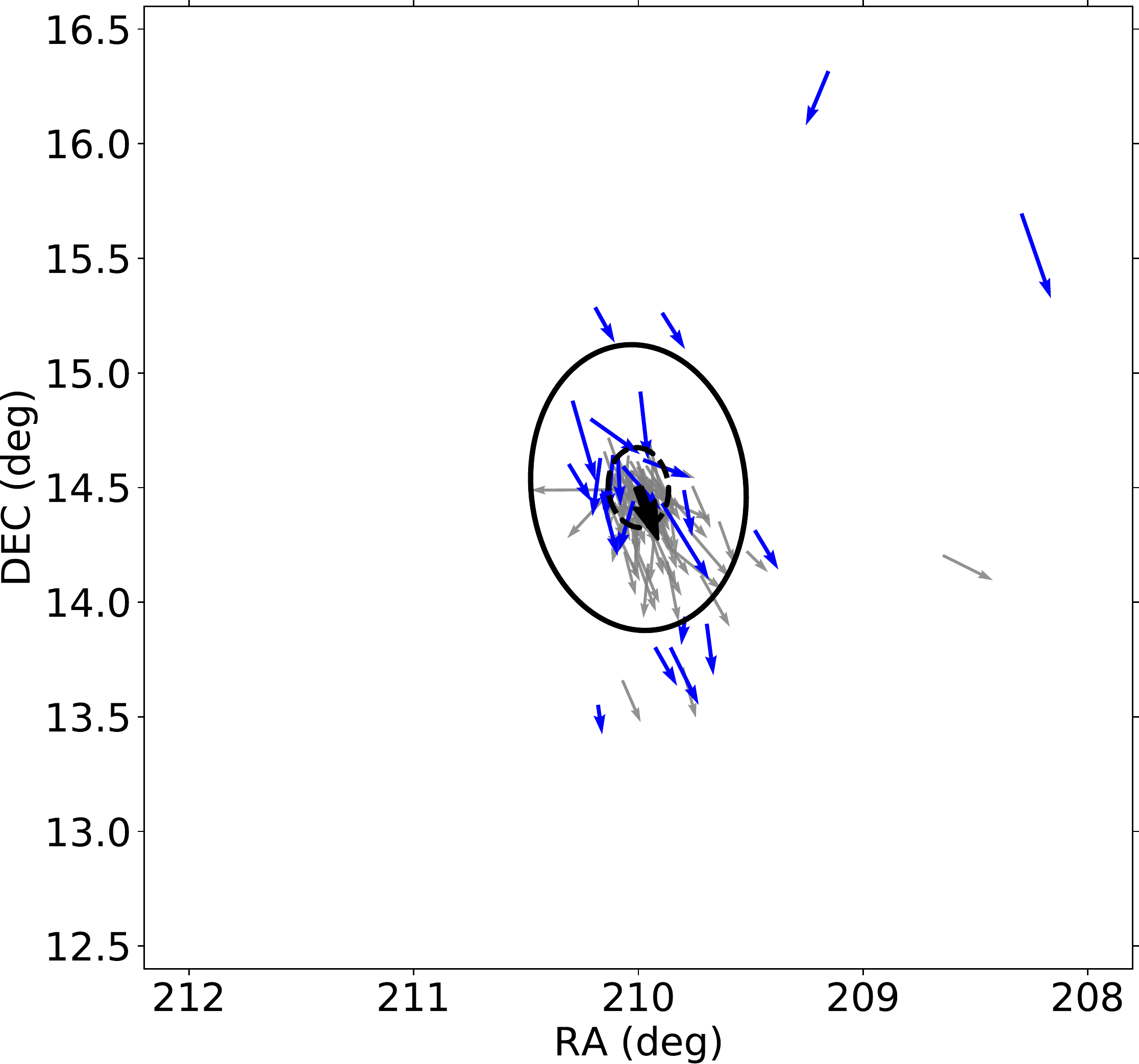}

\caption{{\textit{Gaia}} EDR3 proper motions of previously identified member stars (grey arrows) and of the candidate BHB member stars from this work (blue arrows). The mean proper motion of \boo\  is shown as a thick black arrow (not to scale) and the King profile tidal and half-light ellipses of \cite{munoz} are shown in black solid and dashed lines, respectively. The lengths of the arrows indicating the individual proper motions have been scaled, for clarity, by the magnitude of the mean proper motion, in combination with the sample size .}

\label{fig:mem_pm}
\end{center}
\end{figure}

\begin{figure*}
\begin{center}
\includegraphics[width=\textwidth]{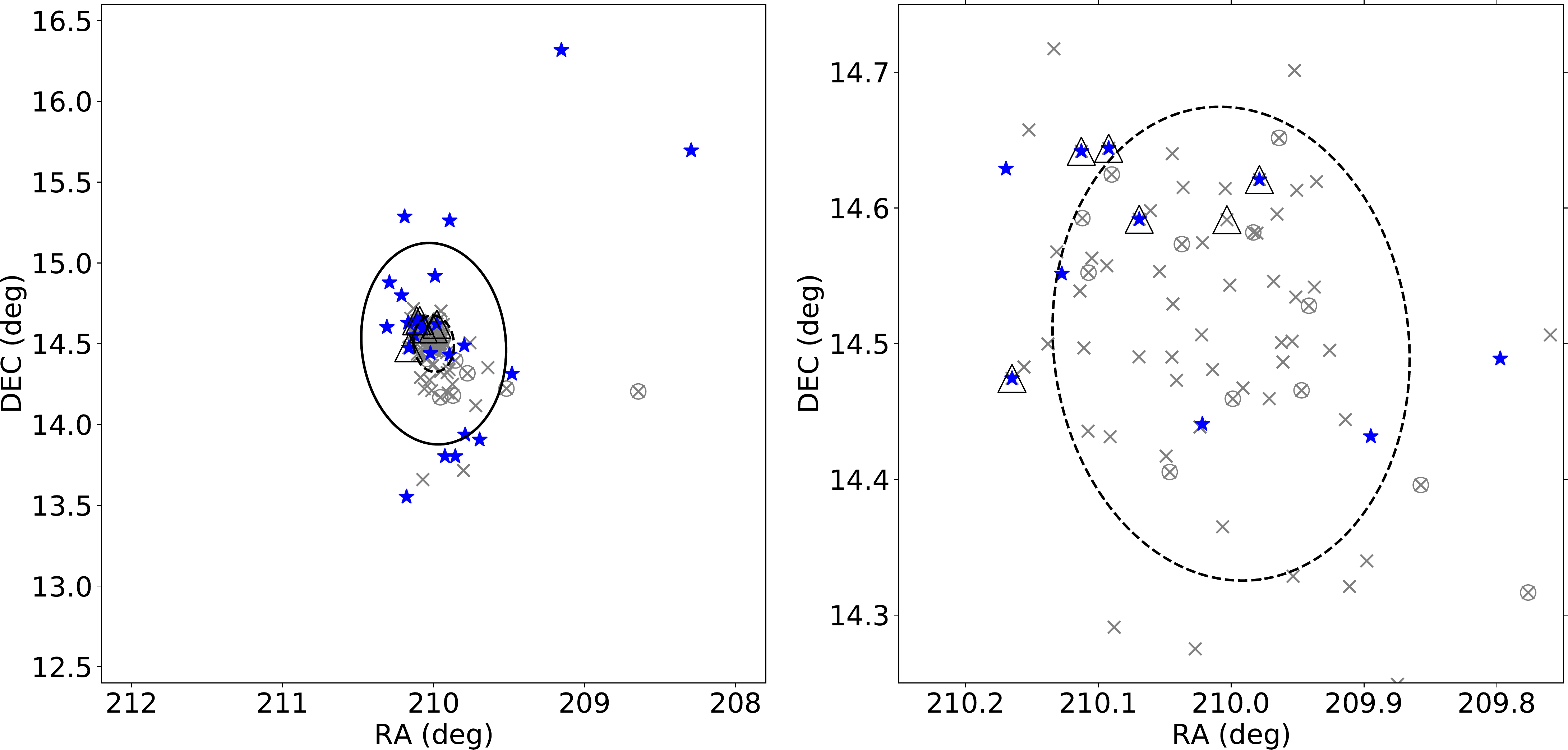}

\caption{The two-dimensional distribution of candidate BHB (blue stars) members of \boo, over-plotted on the King profile tidal and exponential half-light ellipses of \cite{munoz}  (black solid and dashed lines, respectively). Previously identified member stars that have proper motions consistent with \boo\ are shown as grey crosses, open grey circles around crosses indicate that the source is a literature RR Lyrae, and open triangles around the crosses indicate that the source is a non-variable literature BHB (from \citealt{koposov} and \citealt{jenkins}).}

\label{fig:pm_dist}
\end{center}
\end{figure*}

The proper-motion consistent BHB stars identified earlier in this section form our final sample of candidate members among the blue stars (see Table~\ref{tab:blue_off}). Their positions and proper motions are illustrated in Figure~\ref{fig:mem_pm}, with arrows indicating the directions and magnitudes of the individual proper motions (the length of the arrows have been scaled automatically by the mean of the magnitudes of the individual proper motions, in combination with sample size, for clarity). The positions on the sky of these candidate BHB members, together with  previously identified  proper-motion consistent member stars are shown  in Figure~\ref{fig:pm_dist}, with the righthand panel being a zoom-in of the inner regions, where the bulk of the spectroscopic members are located.  The RR Lyrae from \citet{siegel} and \cite{vivas}, in addition to the non-variable BHB stars from \citet{koposov} and \citet{jenkins} are indicated by different symbols. The ellipses in  both Figure~\ref{fig:mem_pm} and Figure~\ref{fig:pm_dist} indicate the half-light radius and King profile tidal radius from \citet{munoz}.

\subsection{Proper-motion Consistent Stars in Off-Fields}

We find between 1 and 4  proper-motion consistent, apparent-magnitude appropriate BHB stars  in each of the off-fields (see Table~\ref{tab:blue_off}). Unlike the \boo\ field, two of the off-fields each contain one apparent-magnitude consistent BSS star with a \textit{Gaia} counterpart, but neither of these stars has a proper motion consistent with that of \boo.

The assumption that these BHB candidates in the off-fields are actually members of the field stellar halo at distances similar to \boo\  and are distributed  approximately uniformly across the sky on $\sim 10^o$ scales implies we could thus expect $\sim 2$ stars in the catalog of candidate BHB members to be non-members, or a $< 10\%$ contamination level even after imposing a proper-motion cut. On the other hand, they could be actual members of \boo\ in a very extended envelope. The location of \boo\ on the sky, in projection against the `Field of Streams', complicates investigations on the structure of \boo\ on these larger angular scales, as illustrated in Appendix~\ref{sec:Sgr}.  In the limit of a smooth and  symmetric stellar halo, the counts of faint BHB in the Milky Way halo should be the same for fields at the same Galactocentric distance, and an alternative approach is to investigate the BHB content of the field halo far from the line-of-sight of \boo, using our technique to classify stars. We adopted Galactocentric Cartesian coordinates, with the Sun located at (X$_\odot$, Y$_\odot$, Z$_\odot$) = ($-8.3$, 0, 0.27)~kpc and \boo\ at (X$_{boo}$, Y$_{boo}$, Z$_{boo}$) = (14.50, $-0.80$, 60.9)~kpc. We selected   a line-of-sight approximately 180$^\circ$ from \boo\ (and thus towards the Galactic anticenter) by changing the sign of the X-coordinate, and sought to identify BHB stars at the same Galactocentric distance as \boo\ (62.6~kpc). We thus applied our classifier to the PS1 data in a $4 \times 4$~degree field centered on Galactic coordinates ($\ell,\,b$) = (187$^\circ$, 84$^\circ$), and restricted the apparent magnitude range to isolate BHB along this line-of-sight with  Galactocentric (X, Y, Z) coordinates of ($-14.5$, $-0.8$, 60.9)~kpc. We found that this field contained fewer BHB stars (27) of any apparent magnitude than did any of the other fields we considered and that there were zero BHB at the same  Galactocentric distance as \boo, allowing for the same selection box as above.  The status of the few BHB in the more local off-fields that have astrometry consistent with being members of \boo\ thus remains ambiguous.

\subsection{Comparison with CMD-based Membership Criteria}

The estimated $\lesssim 10\%$ contamination level of the final proper-motion consistent, KNN-classified sample of BHB associated with \boo\ is significantly lower than  that which would be achieved {\it via\/} selection of stars based on their location on the CMD. For example, if we were to select all blue stars with g-band apparent magnitude within $\pm 0.2$~mag of the  horizontal branch predicted by our chosen PARSEC isochrone, on the $(g_{PS1}-r_{PS1})_0$ vs $g_{PS1_0}$ CMD of the \boo\ field, the resulting sample would contain sixty-three potential BHB members. stars. Application of this criterion to the CMDs of the off-fields would provide samples of between twenty-eight and thirty-six stars. The assumption  that these stars in the off-fields are non-members of \boo\ and that there would be a similar number ($\sim 33$) of non-member stars in the \boo\ field, implies that using only a CMD-based selection cut to identify candidate BHB member stars of \boo\ would be likely to produce a catalog that had $\sim 50\%$ contamination, from primarily Milky Way stars plus background QSOs.

Note that this estimated $\sim 50\%$ contamination rate for CMD-selected candidate BHB is entirely consistent with the fact that all of the candidate member BHB stars selected using this method and observed spectroscopically by \cite{koposov} have radial velocities and proper motions consistent with membership of \boo\ \citep{jenkins}. The stars in this spectroscopic BHB sample are located within or near the half-light radius of \boo, where the contrast between \boo\ members and the field stellar halo is highest (see Figure~\ref{fig:pm_dist}). The expected identification rate  of  non-member BHB candidates based on  the off-fields,  $\sim 33$ per 16~sq.~degree field, would predict between zero and one such contaminant within the area enclosed by the half-light radius of \boo\ ($r_h \simlt 10.5$~arcmin).

\begin{deluxetable}{ccccccccc}
\tablecolumns{2}
\tablewidth{4.0\columnwidth}
\tablecaption{{Candidate Member BHB Per Field} \label{tab:blue_off}}
\tablehead{
\colhead{Field Identifier} &  \colhead{\phm{xxxxxxxxx}}      & \colhead{PM Consistent BHB}
} 

\startdata
\boo & & 24   \\
Off-Field 1 &  &  4 \\
Off-Field 2 & & 1    \\
Off-Field 3 & & 2    \\
Off-Field 4 & & 2   \\
\enddata
\tablecomments{Counts  of apparent-magnitude appropriate BHB stars which also have  proper motion consistent with that of \boo.
}
\end{deluxetable}

\section{Discussion}\label{sec:discussion}

The candidate member stars identified in this work form an extended envelope of blue stars, the two-dimensional distribution of which is presented in Figure~\ref{fig:pm_dist}, together with previously identified members taken from the literature. The King profile tidal and exponential half-light ellipses of \boo\ from \cite{munoz} are indicated for reference. It is evident that the King profile tidal radius does not mark the edge of the member stars, a possibility discussed in \cite{roderick}.  Indeed, our new BHB candidates extend to beyond ten half-light radii from the center of \boo,  out to $\sim 122$ arc minutes, or $\sim 2.3$ kpc. The distribution of these candidate BHB members is generally aligned with the inner isophotes, as suggested by  \citet[see their Fig.~3]{belokurov06} on smaller angular scales than in this analysis, and with their BHB candidates (selected purely on the basis of their location on the CMD) lying predominantly to the North.

We identified  nine candidate BHB members in this extended envelope, three of which are more than one kpc ($\sim 5$ half-light radii) from the center of \boo. Given our low contamination rates, estimated above, we can be confident that these are predominantly  genuine member stars of \boo. Our stars  significantly augment the sample of outer-envelope members of \boo, which previously consisted of only two proper-motion consistent, spectroscopically confirmed RGB member stars (first reported in \citealt{norris}) and two proper-motion consistent RR Lyrae stars \citep{siegel,vivas}. 

After the analysis presented in this work was completed, the complementary investigation of \cite{longeard} was posted to the arXiv. These authors identified candidate \boo\ member stars far from that galaxy's center  using the metallicity-sensitive Pristine photometric filters (see e.g. \citealt{pristine}) in combination with a fiducial isochrone and \textit{Gaia} proper motions. \citet{longeard} obtained  spectroscopic follow-up  of a subset of these stars, all within the 2-degree field-of-view of the 2dF/AAOmega instrument on the Anglo-Australian Telescope (AAT). Their candidate, and confirmed, member stars are RGB and HB stars redder than $(g-i)_{0, SDSS} \sim 0.2$. As such, there is no overlap between the sample of \cite{longeard} and the candidate member stars identified in the present paper. \cite{longeard} found spectroscopically confirmed member stars out to $\sim 4$ half-light radii from the center of \boo\ and thirteen of the 29 total member stars identified in their study reside at projected distances further than two half-light radii from the center of \boo. These redder stars in a radially extended distribution  support our findings from the blue stars.

Returning to the BHB candidate members shown in Figure~\ref{fig:pm_dist}, the human eye is drawn to the string of stars to the south of the King profile tidal ellipse, which, together with the more distant stars to the north, appear to form an S-shape, evocative of tidal arms. Should tidal  distortion and disruption be the cause of  the two-dimensional morphology,  we would expect large-scale tidal debris to be generally aligned with the orbital path of \boo. We investigated this possibility by generating a mock stream of tidal debris from a model satellite galaxy  on an orbit consistent with the present 6-dimensional phase space coordinates of \boo, using the \texttt{MockStreamGenerator} instance within the dynamical analysis package Gala \citep{gala}. While idealized, this exploration should provide insight into the predicted track on the sky of tidal debris, should \boo\ be disrupting.

We defined the stream progenitor by  a Plummer potential \citep{plummer} with mass equal to $10^{7} \mathcal{M}_\odot$ and a scale parameter of 200 pc. These parameter values were chosen to roughly match those derived for \boo\  (see e.g. \citealt{munoz} and Appendix \ref{sec:dynamical}).  For the initial phase-space coordinates of \boo\ we adopted (RA, Dec) of (210.0, 14.5), a heliocentric distance of 65~kpc (e.g. \citealt{okamoto}), mean line-of-sight velocity of 102.5~km~s$^{-1}$ \citep{jenkins}, and the mean proper motion derived in this work. We assumed the \texttt{BovyMWPotential2014} Milky Way potential, which is the Gala implementation of the potential given in \cite{bovy} and described in Appendix \ref{sec:dynamical}. We used the \texttt{FardalStreamDF} class to generate the stream within Gala, such that stream generation would follow the prescription described in \cite{fardal}. We then integrated the orbit backwards in time, and generated the mock stream from the location of the progenitor 500Myr ago, to the present-day location of \boo, in half-million year time-steps. This integration time (0.5~Gyr)  corresponds to approximately a quarter of the (radial) period of \boo\ in this potential, and so can model the effects of only one periGalacticon passage.  Of course, different integration times would produce streams of different lengths. For our present purposes of determining the  general direction of tidal debris that could be generated by \boo\ should it be disrupting, this should suffice.

\begin{figure*}

\begin{center}

\includegraphics[width=.9\textwidth]{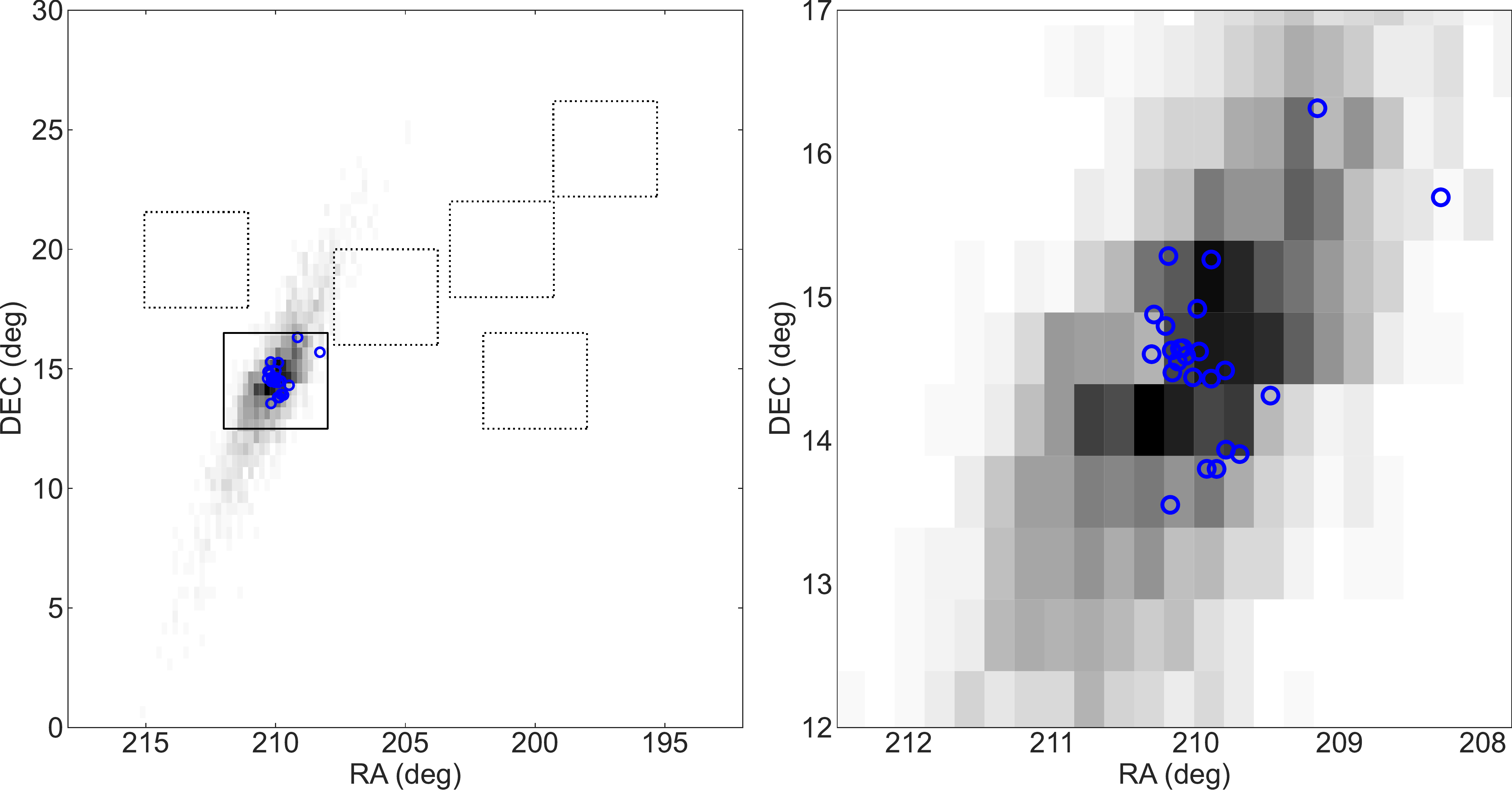}

\caption{The predicted track of a mock stream generated by an idealized model of a satellite galaxy on the same orbit as \boo\ in a fiducial Milky Way potential. The track is shown as a two-dimensional density histogram, where darker grey indicates higher density. It should be noted that the main body of the satellite is not plotted. The candidate BHB member stars of \boo\ are shown as blue  open circles, and the four-by-four degree footprints of the \boo\ field and the off-fields are shown in solid and dotted lines, respectively, in the lefthand panel.}

\label{fig:stream_track}

\end{center}

\end{figure*}

The track of the stream formed in this model is shown in Figure~\ref{fig:stream_track}, together with the candidate BHB member stars found in this work (open blue circles) and the footprints of the four-by-four degree \boo\ field (solid box) and off-fields (dotted boxes). The two-dimensional density histogram of the stream on the sky is shown. The candidate members are, indeed, aligned with the expected orientation of the tidal debris, though are not necessarily coincident with the regions of highest density. Interestingly, it is plausible that off-field one (with center located at (RA, Dec) = (205.75,  18.0)) contains tidal debris from \boo. This possibility is generally consistent with our finding above that this off-field contains a higher number of proper-motion consistent, apparent-magnitude consistent BHB than do the other three off-fields. The detailed morphology of the predicted tidal debris is dependent on the adopted Milky Way potential, the length of time that the progenitor has been disrupting, and the internal parameters of the progenitor and its orbit,  but this exercise indicates that, as a whole, the candidate BHB members of \boo\ are aligned with the track that one would expect should they be tidal debris from \boo. 

The candidate and confirmed member stars of \boo\ presented in \cite{longeard} form an elongated shape similar to that found in this present work, and those authors also explored tidal origins for this morphology using a very  similar approach, and reached consistent conclusions.

Should, alternatively, \boo\ not be tidally  distorted from interacting with the Milky Way, the extended stellar envelope of \boo\ must be produced in some other way. Possible alternative mechanisms for creating an extended stellar envelope invoke mergers or intense stellar feedback to provide the necessary energy injection. Simulations, such as those presented in \cite{tarumi}, show that it is certainly feasible to create an extended stellar distribution in a surviving UFD through an approximately equal-mass merger, after most of the star formation has occurred in the systems. The post-merger system in \cite{tarumi} differs from \boo\ -- it is even more metal poor (mean $[\rm{Fe/H}] \sim -2.7$) and has a radial metallicity gradient that is steeper than that detected in \boo\ (first tentatively discussed in \citealt{norris}, and recently confirmed in \citealt{longeard}). The metallicity gradient in the simulations is enhanced by star formation, and thus chemical enrichment, in the inner regions during (and shortly after) the merger. Purely \lq dry' mergers  after star-formation has ceased could directly deposit stars in the outer regions, as found for one of the dwarf galaxies simulated by \citet{rey}. Placing constraints on the stellar age distribution, in addition to the chemical abundances, is of obvious importance to distinguish these scenarios.

\subsection{ Constraints on the Outer Envelope Stellar Population}\label{sec:expected_bhb}
The presence of nine candidate BHB member stars outside of the King profile nominal tidal radius of \boo\ suggests that there should also be stars in redder phases of stellar evolution at similarly large projected distances. Indeed, as noted above, \citet{norris} and \citet{longeard} identified luminous, evolved red stars with properties consistent with membership of \boo\ far from its center. There should, of course, be many lower-luminosity stars and we estimated their possible contribution to the extended envelope by modeling a synthetic mono-age, mono-metallicity population with a standard IMF. Specifically,  we generated a synthetic population from the PARSEC models \citep{bressan} in the PS1 photometry bands, using the web tool provided at \url{http://stev.oapd.inaf.it/cgi-bin/cmd}. We adopted a total stellar mass of $10^5 \mathcal{M}_\odot$ (chosen purely for simplicity) with a Kroupa broken power-law IMF \citep{kroupa} and assumed  a metallicity of $[\rm{M/H}] = -2.2$ (the lowest metallicity  available) and an age of 13~Gyr.  While not  entirely representative of \boo, this should provide a reasonable estimate for our present  purposes.

We considered a \lq star' in the synthetic population to be a BHB if it had a color $g_{PS1} - r_{PS1} \le 0$ and it was in the core helium-burning phase of evolution, which resulted in 40 BHB \lq stars'. We then checked the number of luminous RGB stars by  selecting  all \lq stars' on the first-ascent red giant branch that would be brighter than $g_{PS1} = 20$ at the distance of \boo, which yielded 44 (bright) RGB \lq stars'. This exercise thus predicts that there should be similar numbers of outer-envelope RGB as BHB, which is generally consistent with the findings of \citet{norris} and \cite{longeard}.

Looking ahead, for example to the Rubin Observatory, we considered all RGB, sub-giant, and main-sequence stars (hereafter \lq redder stars') in the simulated population that would be brighter than $g_{PS1_0} = 25$ if at the distance of \boo\ (thus down to two magnitudes fainter than the MSTO). This predicts  $\sim 135$ redder stars per BHB, and thus $9 \times 135 \sim 1,200$ redder stars (equivalent to $\sim 4,600 \rm{L}_\odot$) in the extended stellar envelope. Assuming a uniform distribution over  a (circular) annulus extending from the King profile nominal tidal radius to the projected radial distance of the furthest candidate member BHB star, i.e.~from $\sim 0.7$ kpc to $\sim 2.3$ kpc, the resulting surface brightness of the (redder) stellar envelope would be $\mu \sim 35$~mag/square~arcsec. Such a low-surface brightness envelope would be difficult to detect, but a suitably clumpy distribution could produce a more tractable over-density on the sky. Indeed, the furthest over-density from the center of \boo\ detected by \cite{roderick} is composed of stars fainter than $g \sim 22.75$. It may thus be possible for future work to leverage deep photometry and identify additional over-densities of redder stars beyond the King profile nominal tidal radius. Finally, we note that the number of candidate BHB member stars that we have identified beyond the areal coverage of earlier surveys implies that \boo\ may have a higher total luminosity than previously estimated, a possibility that is also suggested by the relatively large number of associated RR Lyrae \citep{siegel}.

\section{Conclusion}\label{sec:conclusion}

This work explored the outer extents of the \boo\ UFD, and revealed a stellar envelope of evolved blue stars that is more extended than previously thought. We investigated BHB and BSS stars as tracers of  the stellar population of \boo, as these blue stars occupy a distinct region in CMD-space that has relatively low contamination from either Milky Way stars and stellar remnants or extragalactic sources. We used a k-nearest neighbors classifier to identify likely BHB and BSS stars in Pan-STARRS \textit{griz} photometry, as various color combinations of these filters has been shown to be sensitive to surface gravity in blue stars (see e.g. \citealt{vickers}). We found twenty five BHB and four BSS within a $4 \times 4$ degree field centered on \boo\  at appropriate apparent magnitudes to be members. We then cross-matched with \textit{Gaia} EDR3, and determined that twenty-four of these twenty-five BHB have proper motions consistent with \boo. None of the candidate member BSS stars has high-quality \textit{Gaia} information and thus we could not quantify their proper motions.  We then determined the expected contamination from non-member sources in the final sample of candidate \boo\ members by repeating this selection process in various off-fields. From this exercise, we determined that the final \boo\ candidate BHB member sample has a likely  contamination level of $\simlt 10\%$. 

Nine of the twenty-four candidate BHB member stars are at projected radii beyond the King profile nominal tidal radius derived by \cite{munoz}, more than doubling the known population of member stars in these outer regions, including those of \citet{longeard}. These outer envelope stars confirm that \boo\ has an extended spatial distribution, as indicated in e.g.~\cite{roderick}. Their 2-dimensional morphology  is evocative of that produced by tidal interactions. We explored this possibility by using the \texttt{MockStreamGenerator} within Gala to generate a mock stellar stream originating from an idealized satellite-galaxy progenitor, with properties similar to those inferred for \boo. The candidate BHB member stars are, indeed, aligned with the resulting mock stream, indicating that tidal interactions are a plausible explanation for the observed morphology.

Candidate member stars at large distances from the center of their host galaxy, such as those identified here, present excellent targets for wide-field spectroscopic surveys. These stars present exciting opportunities to constrain the formation and evolution of their host systems. Wide-field spectroscopic instruments, such as the 2dF/AAOmage on the AAT and the Prime Focus Spectrograph on the Subaru telescope in Mauna Kea, will be able to observe stars over many square degrees in a small number of pointings, which will enable the exploration of the furthest extents of nearby dwarf and ultra-faint dwarf galaxies. 

\section{ Acknowledgments}

CF and RFGW acknowledge support through the generosity of Eric and Wendy Schmidt, by recommendation of the Schmidt Futures program. CF and RFGW thank the anonymous referee for their thorough report. RFGW thanks her sister, Katherine Barber, for her support  of this research. CF and RFGW thank Imants Platais and Vera Kozhurina-Platais for insightful discussions.

The Pan-STARRS1 Surveys (PS1) and the PS1 public science archive have been made possible through contributions by the Institute for Astronomy, the University of Hawaii, the Pan-STARRS Project Office, the Max-Planck Society and its participating institutes, the Max Planck Institute for Astronomy, Heidelberg and the Max Planck Institute for Extraterrestrial Physics, Garching, The Johns Hopkins University, Durham University, the University of Edinburgh, the Queen's University Belfast, the Harvard-Smithsonian Center for Astrophysics, the Las Cumbres Observatory Global Telescope Network Incorporated, the National Central University of Taiwan, the Space Telescope Science Institute, the National Aeronautics and Space Administration under Grant No. NNX08AR22G issued through the Planetary Science Division of the NASA Science Mission Directorate, the National Science Foundation Grant No. AST-1238877, the University of Maryland, Eotvos Lorand University (ELTE), the Los Alamos National Laboratory, and the Gordon and Betty Moore Foundation.

Funding for SDSS-III has been provided by the Alfred P. Sloan Foundation, the Participating Institutions, the National Science Foundation, and the U.S. Department of Energy Office of Science. The SDSS-III web site is http://www.sdss3.org/. SDSS-III is managed by the Astrophysical Research Consortium for the Participating Institutions of the SDSS-III Collaboration including the University of Arizona, the Brazilian Participation Group, Brookhaven National Laboratory, Carnegie Mellon University, University of Florida, the French Participation Group, the German Participation Group, Harvard University, the Instituto de Astrofisica de Canarias, the Michigan State/Notre Dame/JINA Participation Group, Johns Hopkins University, Lawrence Berkeley National Laboratory, Max Planck Institute for Astrophysics, Max Planck Institute for Extraterrestrial Physics, New Mexico State University, New York University, Ohio State University, Pennsylvania State University, University of Portsmouth, Princeton University, the Spanish Participation Group, University of Tokyo, University of Utah, Vanderbilt University, University of Virginia, University of Washington, and Yale University.

This work has made use of data from the European Space Agency (ESA) mission
{\it Gaia} (\url{https://www.cosmos.esa.int/gaia}), processed by the {\it Gaia}
Data Processing and Analysis Consortium (DPAC,
\url{https://www.cosmos.esa.int/web/gaia/dpac/consortium}). Funding for the DPAC
has been provided by national institutions, in particular the institutions
participating in the {\it Gaia} Multilateral Agreement.

This work made use of data from the Guoshoujing Telescope (the Large Sky Area Multi-Object Fiber Spectroscopic Telescope LAMOST),  a National Major Scientific Project built by the Chinese Academy of Sciences. Funding for the project has been provided by the National Development and Reform Commission. LAMOST is operated and managed by the National Astronomical Observatories, Chinese Academy of Sciences.

\software{Astropy \citep{astropy_2013, astropy_2018}, dustmaps \citep{green}, Gala \citep{gala}, Galpy \citep{bovy}, scikit-learn \citep{sklearn}, TOPCAT \citep{topcat}}

\appendix

\section{ Details of the Testing and Implementation of the KNN Classifier}\label{sec:knn_appendix}

The  fractions of sources  of the different types (BHB, BSS/MS, WD, and QSOs) in the SEGUE-PS1 catalog are not equal, which can bias the classifier (see e.g.~\citealt{patel} for a discussion of the effect of class imbalance on  KNN classification). There are, in total,  4995 BSS/MS stars, 2875 BHB stars, 475 WDs, and 280 QSOs with $-0.5\le (g_{PS1}-r_{PS1})_0 \le 0$ in the catalog,  so  that more than half of the catalog is BSS/MS stars,  while approximately a third is BHB stars. Class imbalance generally causes a classifier to be baised towards labelling  objects as members of the majority class,  which in our case is  BSS stars. We mitigated this by resampling the SEGUE-PS1 data to create a balanced dataset for training and testing, as discussed below.

We opted to create a resampled catalog where each class (BSS/MS, BHB, WD, QSO)  had approximately the same number of sources as do the BHBs in the original, SEGUE-PS1  catalog. This produced resampled catalogs containing $\sim 11,500$ sources.
We under-sampled the BSS/MS class, i.e.~we took a random subset of the stars in the BSS/MS class,  while  over-sampling the WD and QSO classes. We generated the required additional synthetic WD and QSO sources by first randomly selecting a WD or QSO source (with replacement) from the original SEGUE-PS1 catalog  and then drawing 
\textit{griz} magnitudes  -- to be assigned to the  new synthetic source -- from Gaussian distributions  defined by the means and uncertainties of the PS1 photometry for the original source.  We  did explore another approach to over-sampling the QSO and WD samples (further discussed below), and opted to employ the approach just described, based on  both its simplicity and the fact that it  avoided  the use of additional external data. 

We used the \texttt{KNeighborsClassifier} module implemented in the  scikit-learn Python package, adopting a Minkowski distance metric, and weighted the votes of each neighbor by  the inverse of its distance  from the input object  such that closer training points have higher weight in deciding the label for the input object. We constructed two  identically indexed arrays from the resampled SEGUE-PS1 catalog, such that  every  unique source had the same index in both arrays. One of the arrays contained every possible color combination of the PS1 colors for each source, while the other array contained the label for that source. We then randomly split these arrays into a training set, containing $70\%$ of the  sample, and a test set,  which contained the remaining $30\%$.  We produced  ten different resampled SEGUE-PS1 catalogs,  each with its  associated training and test sets, to reduce the dependence of the results on any one individual realization of the under-sampled BSS/MS stars or over-sampled WDs and QSOs. 

We then determined the optimal k-value for the KNN classifier by training and testing  it using a range of k-values between ${\rm k=1}$ and ${\rm k=100}$.  We found that the completeness and purity of each class { of source (as defined in Section~\ref{sec:knn-main} of the main text) varied with the different k-values, such that smaller k-values $(\le 10)$ produced  lower overall values for the completeness and purity.  We found that k-values of around 20 produced samples of high completeness and purity for each class and the performance reached a plateau beyond that,   motivating our choice of $k=20$. 

With the optimal value of k identified, we then trained a KNN classifier on each of the training sets (i.e.~we trained ten classifiers, each with the same parameters but different training data). We then applied the classifiers to the test sets and computed the completeness and purity of each output source type, or class. We also computed the confusion matrix for each pair of training set and test set, with an example shown in  Table~\ref{tab:conf_mat}. Each row of the confusion matrix indicates the true class, and each column indicates the class that was assigned by the classifier. Thus entries on the diagonal of the matrix represent correctly labeled sources,  while the off-diagonal entries  represent mis-classifications. For example, the first row of the confusion matrix in Table~\ref{tab:conf_mat} indicates that 728 BHB were correctly classified, 103 BHB  were classified as BSS, 12 BHBs  were classified as WDs, and 20 BHBs were classified as QSOs.   The first column of the matrix in Table~\ref{tab:conf_mat} indicates that 144 BSS  were classified as BHB, 34 WDs  were classified as BHB, and 4 QSO  were classified as BHB. The last two rows give the overall purity and completeness for each class in this resampled SEGUE-PS1  catalog.  The final  purity and completeness for each class was obtained by  averaging over all the resampled datasets and is reported in the first two rows of Table~\ref{tab:knn_comp}.

 \begin{deluxetable}{ccccccccc}
\tablecolumns{9}
\tablewidth{0pt}
\tablecaption{{ Confusion Matrix} \label{tab:conf_mat}}
\tablehead{
\colhead{} &   \colhead{\phm{xxxxxxxxx}} &  \colhead{BHB} &  \colhead{\phm{xxxxxxxxx}} & \colhead{BSS} &  \colhead{\phm{xxxxxxxxx}} & \colhead{WD} &  \colhead{\phm{xxxxxxxxx}} & \colhead{QSO}
} 

\startdata
BHB & & 728 & & 103 & &  12 & & 20   \\
BSS & & 144 & & 692  & & 16 & & 12 \\
WD & & 34  & & 79 & & 732 & & 12   \\
QSO &  & 4 & &   19 &  & 25 & &  805  \\
Purity &  & 80$\%$ &  & 77$\%$ &  & 93$\%$ & & 95$\%$ \\
Completeness & &  84$\%$ &  & 80$\%$ &  & 85$\%$ &  & 94$\%$ \\
\enddata
\tablecomments{A representative confusion matrix for the KNN classifier resulting from training and testing on a random resampling of the SEGUE-PS1 catalog, as described in the text, with the training and test subsets containing, respectively, $70\%$ and $30\%$ of the sources.  Each row in the table indicates the true label, and each column indicates the assigned label, such that the  entries along the diagonal give the number of sources that  were given correct labels, and off-diagonal entries  show the number of  mis-classifications into each other label. The purity and completeness of each type of source  are given in the last two rows.
}
\end{deluxetable}

It is evident, from inspection of the confusion matrix,  that it is rare for QSO sources to be mis-classified as either BHB or BSS/MS. This may have been anticipated, as QSOs are generally well-separated from the loci of BHB and BSS/MS in color-space (see e.g. Figure~\ref{fig:segue_ccd}).  The false-positive contribution of WDs to the BHB class is smaller than that  of WDs to the BSS/MS class, which is perhaps unsurprising given that BHB stars have the lowest surface gravities out of the three and should thus be more well-separated from the highest surface gravity sources in color-space. The WD and QSO classes have the highest purity of the four classes, each with greater than $90\%$ purity, approximately $10\%$ higher than  those of the BHB and BSS/MS classes. 
 
We investigated  how  the performance of the classifier  depended on the  approach taken to over-sampling the WD and QSO classes. In this comparison, instead of producing synthetic WD and QSOs, we cross-matched external WD and QSO catalogs  with  the PS1 database and added a random subset of these additional sources to the SEGUE-PS1  catalog, again creating a dataset with equal  sample sizes in each class of source and $\sim 11,500$ sources  in total. We adopted the WD catalog of \cite{wd} and considered only those sources sources that have  both a high probability of being a white dwarf (\texttt{Pwd }$> .95$)  and a small uncertainty in that probability (\texttt{f\_Pwd} $< 1$). We adopted the QSO catalog of \cite{qso}. We employed the same matching and photometric quality criteria as described above for the SEGUE-PS1 catalog, and further required that the cross-matched sources have magnitudes within the $g_{PS1}$ magnitude range of the SEGUE-PS1 sample. We then randomly selected subsets of these data to add to the SEGUE-PS1 catalog, and again randomly under-sample the BSS class, and re-trained and re-tested the classifiers on ten random iterations of these new, augmented data. 

 \begin{deluxetable}{ccccccccc}
\tablecolumns{9}
\tablewidth{0pt}
\tablecaption{ Average Purity and Completeness \label{tab:knn_comp}}
\tablehead{
\colhead{ } &   \colhead{\phm{xxxx}} &  \colhead{BHB} &   \colhead{\phm{xxxx}} & \colhead{BSS} &   \colhead{\phm{xxxx}} & \colhead{WD} &   \colhead{\phm{xxxx}} & \colhead{QSO}
} 

\startdata
Purity, Synthetic Sources & & 81$\%$ & & 80$\%$ & & 95$\%$ & & 95$\%$ \\
Completeness, Synthetic Sources & &  82$\%$ & &  78$\%$ & &  84$\%$ & &  95$\%$ \\
Purity, External Sources &  & 81$\%$ &  & 78$\%$ &  & 94$\%$ & &  95$\%$ \\
Completeness, External Sources &  & 82$\%$ &  & 76$\%$ &  & 84$\%$ &  &  95$\%$ \\
\enddata
\tablecomments{The average purity and completeness of each class using the output of the classifiers trained on the a re-balanced catalog created with first synthetic WDs and QSOs (upper two rows) and second the addition of WDs and QSOs from external catalogs  (lower two rows). In all cases, the probability threshold is $50\%$, i.e. sources must be assigned a given label with  higher than $50\%$ probability in order to be classified  as that type. } 
\end{deluxetable}

The resultant  average purity and completeness of each class using the classifiers trained on these new data, created by including external catalogs, are given in the bottom two rows of Table~\ref{tab:knn_comp}.  There is evidently little difference between the results obtained from either approach to rebalancing the different classes in the resampled SEGUE-PS1 catalog.  As such, we proceeded with the first approach, in which we created synthetic sources.

We  note, for completeness, that \cite{smith} also investigated the use of a KNN classifier for the identification of BHB stars in broad-band photometric catalogs. However,  their approach differs from that taken here  in a few fundamental ways, which makes direct comparison of the classification results difficult. For example, \cite{smith} use four color combinations of the SDSS ugri bands, and label sources in a binary fashion, simply as either  BHB or not-BHB. Further, they use an equal ratio of BHB to not-BHB sources in the training set, but they adjust the output label probabilities with assumed priors. They do, however, provide the raw completeness,  obtained before application of these priors, for the BHB class from their KNN classifier and their quoted value of $81.1\%$ is approximately the same as that found by  our classifier.

\section{Dynamical Analysis}\label{sec:dynamical}
We derived the orbit of \boo\ in two different Milky Way gravitational potentials (described below), which we use to set limits on the mass of \boo. We adopt, as initial conditions for the orbit integrations,  the mean line-of-sight velocity of \boo\ given in \cite{jenkins}, the heliocentric distance estimate of \cite{okamoto}, and the mean proper motion we derived above. We performed a thousand Monte Carlo realizations of the orbit in each potential (assumed static), integrating each orbit over the time interval $-5$~Gyr to $+10$~Gyr. In each Monte Carlo realization, we drew the values of the component of proper motion ($\mu_{\alpha *}$, $\mu_{\delta}$), distance, and line-of-sight velocity from Gaussian distributions defined by the quoted means and errors of each parameter.  The median and $16^{th}$ and $84^{th}$ percentiles of the resulting distribution of  key orbital parameters for each adopted Milky Way potential are given in Table~\ref{tab:orbits}. 

 We first adopted  the potential \texttt{MWPotential2014}, as implemented in \texttt{galpy} \citep[described in detail in][]{bovy}. The mass distribution corresponding to this potential has a virial mass of $0.8 \times 10^{12} \mathcal{M}_\odot$ and is an idealized model consisting of a Navarro-Frenk-White (NFW) dark-matter halo \citep{nfw}, a spherical bulge, and a Miyamoto-Nagai disk \citep{miyamoto}. For comparison, we also adopted the \texttt{McMillan17} potential, again as implemented in \texttt{galpy} \citep[described in detail in][]{mcmillan}. The mass distribution of this second potential has a total virial mass of $1.3 \times 10^{12} \mathcal{M}_\odot$, and is a more complex model fitted to the observational data,  again assuming an NFW halo. As noted, in both cases, the Milky Way potential is static, and we neglected any contributions from  other systems (such as the  Large Magellanic Cloud). 

\begin{deluxetable}{ccccccccc}
\tablecolumns{5}
\tablewidth{0pt}
\tablecaption{ Derived Orbital Parameters \label{tab:orbits}}
\tablehead{\colhead{} &   \colhead{\phm{xxxxxx}} &  
\colhead{MWPotential2014} &   \colhead{\phm{xxxxxx}} &  \colhead{McMillan17} 
} 

\startdata
Apocenter (kpc) & & $105.36^{+22.27}_{-15.68}$ & & $89.03^{+13.65}_{-9.27}$   \\
Pericenter (kpc) & & $42.73^{+5.88}_{-6.71}$ & &   $37.44^{+6.49}_{-5.85}$  \\
Eccentricity & & $0.43^{+0.02}_{-0.01}$ & & $0.41^{+0.02}_{-0.01}$   \\
\enddata
\tablecomments{Median orbital parameters of \boo, from 1000 Monte Carlo realizations of the initial conditions.}
\end{deluxetable}

Under the assumption that all of the BHB candidate member stars are gravitationally bound to \boo,  their maximum radial distance from the center of \boo\ (2.3~kpc in projection) can be used as a estimate of the tidal limit.  We estimated the mass of \boo\ in the two different Milky Way potentials using the \texttt{rtide} function within \texttt{galpy} \citep[following][]{chiti}. The \texttt{rtide} function assumes a point-mass potential for the satellite and that it is on a circular orbit.  In the lower mass \texttt{MWPotential2014} potential, a tidal radius of $r_t = 2.3$ kpc requires that \boo\ have a mass of $6.4 \times 10^7 \mathcal{M}_\odot$. In the higher mass \texttt{McMillan17} potential, the same tidal radius requires that \boo\ have a mass of $10.8 \times 10^7 \mathcal{M}_\odot$. For comparison, the King profile nominal tidal radius ($r_t = 710$ pc) requires that \boo\ have a mass of $0.2 \times 10^7 \mathcal{M}_\odot$ in the \texttt{MWPotential2014} potential and $0.3 \times 10^7 \mathcal{M}_\odot$ in the \texttt{McMillan17} potential,  while the dynamical mass within the half-light radius has been estimated from the internal velocity dispersion to be $\sim 0.5 \times 10^7 \mathcal{M}_\odot$ \citep{jenkins}.

Adopting, instead, the expression for the tidal radius of a (point mass) cluster on an elliptical orbit around a (point mass)  galaxy from \cite{king} (his Equation~11), gives very similar results. In this approximation, we estimated the mass interior to the orbit of \boo\ in the \texttt{MWPotential2014} model of the Milky Way using the \texttt{mass\_enclosed} method within Gala, which assumes that the potential is spherical. Adopting the ellipticity from Table~\ref{tab:orbits} and  rearranging Equation~11 of \cite{king}, we find that a tidal radius of $r_t = 2.3$ kpc requires that \boo\  have a mass of $22.2 \times 10^7 \mathcal{M}_\odot$. Adopting instead the King profile nominal tidal radius for \boo\ from \cite{munoz}, $r_t = 710$ pc, would require \boo\ to have a mass of $0.65 \times 10^7 \mathcal{M}_\odot$. The validity of the assumption that all stars are gravitationally bound to \boo\ is yet unknown, but these exercises indicate that \boo\ may be significantly more massive than estimated from data primarily within the central regions. We note that the higher values we obtained for the total mass, $\sim 10^8$~M$_\odot$, are entirely consistent with the virial masses of UFDs obtained by \citet{nadler} from simulations of the Milky Way, together with its satellites.

\section{Additional, Sagittarius Stream Off-field}\label{sec:Sgr}

In this section  we discuss an additional $4^\circ \times 4^\circ$ off-field, centered on (RA, Dec) = (200, 14.5), that we rejected due to the  clear signal of the Sagittarius (Sgr) Stream. The proper motions of stellar sources in this field are  shown in Figure~\ref{fig:off_five}; the prominent  over-density at $(\mu_{\delta*}, \mu_{\alpha}) \sim (-1.1, -0.75)$ mas yr$^{-1}$ is most likely due to stars from the Sgr Stream. This field has a larger number of blue star-like sources (those with color $-0.5 \le (g_{PS1} - r_{PS1})_0 \le 0$) than any other field investigated in this work, with a total of 455 blue sources. We applied our KNN Classifier to these data and identified large numbers of both BHB and BSS stars (81 BHB and and 67 BSS in total). The enhanced presence of BHB and BSS, compared to the remaining fields,  is consistent with the presence of the Sgr Stream. Further, only three of the identified BHB stars in this field have appropriate apparent magnitudes to be at the distance of \boo. Following the procedure described in the main text, we cross-matched the PS1 catalog for this field with \textit{Gaia} EDR3, and identified stars that have proper motions consistent with that of \boo. Only one of the BHB stars at an appropriate apparent magnitude also has a proper motion that is consistent with that of \boo. This low number is in agreement with the results reported above for the other off-fields. We can thus conclude that any possible contamination from the Sgr Stream in the \boo\ candidate member BHB sample is minimal.

\begin{figure}
\begin{center}

\includegraphics[width=.5\textwidth]{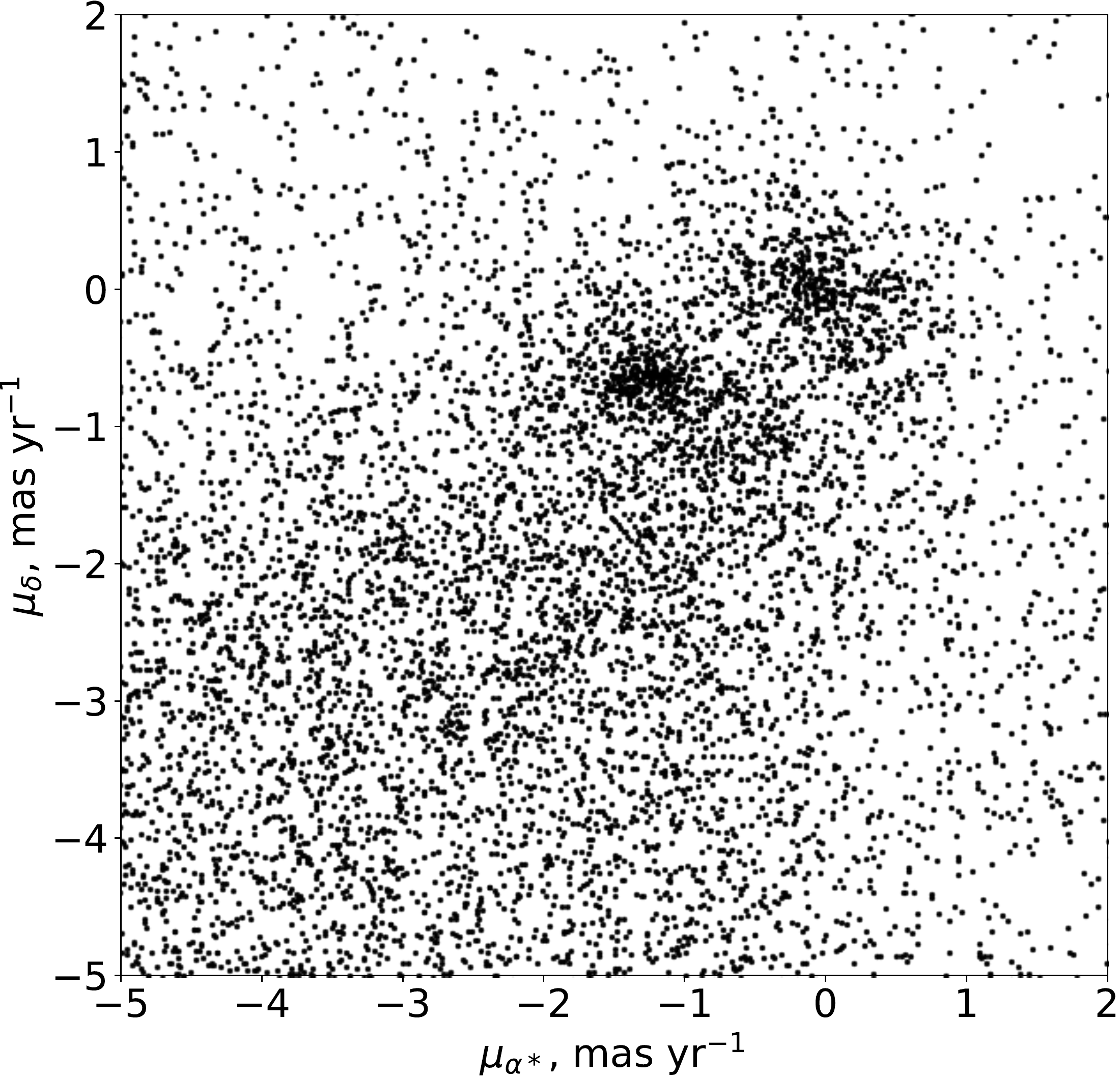}

\caption{The \textit{Gaia} EDR3 proper motions of sources in the rejected off-field. The over-density centered on $\mu_{\delta*} = \mu_{\alpha} = 0$ is present in all fields used in this work, and is likely due to very distant sources, such as AGN and QSOs. The prominent over-density at $(\mu_{\delta*}, \mu_{\alpha}) \sim (-1.1, -0.75)$ mas yr$^{-1}$ is most likely due to stars in the Sagittarius Stream and is not present in the other fields.}

\label{fig:off_five}

\end{center}
\end{figure}


\end{document}